\documentclass[letterpaper,12pt]{article}

\usepackage{graphicx} 

\usepackage{amssymb,amsmath,aastex_hack}
\usepackage[margin=1.25in,includehead]{geometry}

\newcommand{\rearth}{\mbox{$R_{\ensuremath{\oplus}}$}}
\newcommand{\mearth}{\mbox{$M_{\ensuremath{\oplus}}$}}
\def\sun{\odot}
\newcommand{\numax}{\mbox{$\nu_{\rm max}$}}
\newcommand{\Dnu}{\mbox{$\Delta \nu$}}

\newcommand{\muHz}{\mbox{$\mu$Hz}}
\newcommand{\teff}{\mbox{$T_{\rm eff}$}}
\newcommand{\logg}{\mbox{$\log g$}}

\newcommand{\mh}{\mbox{$\rm{[M/H]}$}}
\newcommand{\msun}{\mbox{$M_{\sun}$}}
\newcommand{\rsun}{\mbox{$R_{\sun}$}}
\usepackage{url}

\usepackage{scicite}
\newenvironment{sciabstract}{%
\begin{quote} \bf}
{\end{quote}}

\usepackage{times}

\newcommand{\id}{\mbox{Kepler-56}}

\title{\textbf{Stellar Spin-Orbit Misalignment in a Multiplanet System}}

\date{}

\def\apjl{ApJ}%
\def\apjs{ApJS}%
\def\ao{Appl.~Opt.}%
\def\aap{A\&A}%
\begin{document} 

\begin{center}
\LARGE{\textbf{Stellar Spin-Orbit Misalignment in a \makebox{Multiplanet System}}}
\vspace{0.1cm}
\end{center}

\begin{center}
\begin{large}
\makebox{Daniel Huber$^{1,\ast}$}, 
\makebox{Joshua A.\ Carter$^{2}$}, 
\makebox{Mauro Barbieri$^{3}$}, 
\makebox{Andrea Miglio$^{4,5}$},  
\makebox{Katherine M.\ Deck$^{6}$}, 
\makebox{Daniel C.\ Fabrycky$^{7}$}, 
\makebox{Benjamin T.\ Montet$^{8}$}, 
\makebox{Lars A.\ Buchhave$^{9,10}$}, 
\makebox{William J.\ Chaplin$^{4,5}$}, 
\makebox{Saskia Hekker$^{11,12}$}, 
\makebox{Josefina Montalb\'{a}n$^{13}$}, 
\makebox{Roberto Sanchis-Ojeda$^{6}$}, 
\makebox{Sarbani Basu$^{14}$}, 
\makebox{Timothy R. Bedding$^{15,5}$}, 
\makebox{Tiago L. Campante$^{4,5}$}, 
\makebox{J{\o}rgen Christensen-Dalsgaard$^{5}$}, 
\makebox{Yvonne P. Elsworth$^{4,5}$}, 
\makebox{Dennis Stello$^{15,5}$}, 
\makebox{Torben Arentoft$^{5}$}, 
\makebox{Eric B.\ Ford$^{16,17}$}, 
\makebox{Ronald L.\ Gilliland$^{16}$}, 
\makebox{Rasmus Handberg$^{4,5}$}, 
\makebox{Andrew W.\ Howard$^{18}$}, 
\makebox{Howard Isaacson$^{19}$}, 
\makebox{John Asher Johnson$^{8}$},  
\makebox{Christoffer Karoff$^{5}$}, 
\makebox{Steven D.\ Kawaler$^{20}$}, 
\makebox{Hans Kjeldsen$^{5}$}, 
\makebox{David W.\ Latham$^{2}$},
\makebox{Mikkel N.\ Lund$^{5}$}, 
\makebox{Mia Lundkvist$^{5}$},
\makebox{Geoffrey W.\ Marcy$^{19}$}, 
\makebox{Travis S.\ Metcalfe$^{21,5}$}, 
\makebox{Victor Silva Aguirre$^{5}$} and
\makebox{Joshua N.\ Winn$^{6}$}
\end{large}
\end{center}

\begin{center}
\footnotesize{$^{1}$NASA Ames Research Center, MS 244-30, Moffett Field, CA 94035, USA}\\
\footnotesize{$^{2}$Harvard-Smithsonian Center for Astrophysics, 60 Garden Street, Cambridge, MA 02138, USA}\\
\footnotesize{$^{3}$CISAS, University of Padua, via Venezia 15, 35131 Padova, Italy} \\
\footnotesize{$^{4}$School of Physics and Astronomy, University of Birmingham, Birmingham B15 2TT, UK}\\
\footnotesize{$^{5}$}{Stellar Astrophysics Centre, Department of Physics and Astronomy, Aarhus University,} \\
\footnotesize{Ny Munkegade 120, 8000 Aarhus C, Denmark} \\
\footnotesize{$^{6}$Department of Physics, Massachusetts Institute of Technology, 77 Massachusetts Ave.,}\\
{Cambridge, MA 02139, USA} \\
\footnotesize{$^{7}$Department of Astronomy and Astrophysics, University of Chicago, 5640 S. Ellis Ave.,} \\ 
{Chicago, IL 60637, USA} \\
\footnotesize{$^{8}$Department of Astrophysics, California Institute of Technology, MC 249-17, Pasadena, CA 91125} \\
\footnotesize{$^{9}$Niels Bohr Institute, University of Copenhagen, DK-2100 Copenhagen, Denmark} \\
\footnotesize{$^{10}$Centre for Star and Planet Formation, Natural History Museum of Denmark,} \\ 
\footnotesize{University of Copenhagen, DK-1350 Copenhagen, Denmark} \\
\footnotesize{$^{11}$Astronomical Institute ``Anton Pannekoek'', University of Amsterdam,} \\ 
{1098 XH Amsterdam, The Netherlands} \\
\footnotesize{$^{12}$Max-Planck-Institut f\"ur Sonnensystemforschung, 37191 Katlenburg-Lindau, Germany} \\ 
\footnotesize{$^{13}$Institut d'Astrophysique et de G\'{e}ophysique de l'Universit\'{e} de Li\`{e}ge, All\'ee du 6 Ao\^{u}t 17,} \\
{B 4000 Li\`{e}ge, Belgium} \\
\footnotesize{$^{14}$ Department of Astronomy, Yale University, PO Box 208101, New Haven, CT 06520-8101} \\
\footnotesize{$^{15}$ Sydney Institute for Astronomy (SIfA), School of Physics, University of Sydney, NSW 2006, Australia} \\
\footnotesize{$^{16}$Center for Exoplanets and Habitable Worlds, The Pennsylvania State University,} \\
{University Park, PA 16802} \\
\footnotesize{$^{17}$Astronomy Department, University of Florida, 211 Bryant Space Sciences Center,} \\
{Gainesville, FL 32111, USA} \\
\footnotesize{$^{18}$Institute for Astronomy, University of Hawaii, 2680 Woodlawn Drive, Honolulu, HI 96822, USA} \\
\footnotesize{$^{19}$Department of Astronomy, University of California, Berkeley, CA 94720, USA} \\
\footnotesize{$^{20}$Department of Physics and Astronomy, Iowa State University, Ames, IA 50011 USA} \\
\footnotesize{$^{21}$Space Science Institute, Boulder, CO 80301, USA} \\
\footnotesize{$^\ast$To whom correspondence should be addressed; E-mail:  daniel.huber@nasa.gov}
\end{center}


\baselineskip24pt

\begin{sciabstract}
{Stars hosting hot Jupiters are often observed to have high 
obliquities, whereas stars with multiple co-planar 
planets have been seen to have low obliquities. This 
has been interpreted as evidence that hot-Jupiter 
formation is linked to dynamical disruption, 
as opposed to planet migration through a protoplanetary disk.
We used asteroseismology to measure 
a large obliquity for \id, a red giant star hosting two transiting co-planar 
planets. These observations show that spin-orbit misalignments are 
not confined to hot-Jupiter systems. Misalignments in a broader class of systems
had been predicted as a consequence of torques from wide-orbiting 
companions, and indeed radial-velocity measurements revealed a third companion in a 
wide orbit in the \id\ system.}
\end{sciabstract}

The Kepler space telescope detects exoplanets by measuring 
periodic dimmings of light as a planet passes in front of its host star\cite{borucki10}. 
The majority of the $\sim$\,150,000 targets observed by 
Kepler are unevolved stars near the main sequence, because those stars provide the 
best prospect for detecting habitable planets similar to Earth\cite{batalha10}. 
In contrast, the temperature and surface gravity of 
\id\ (KIC\,6448890) indicate that it is an evolved star with exhausted hydrogen in its core, 
and that it started burning hydrogen in a shell surrounding an inert Helium core. 
Stellar evolutionary theory 
predicts that our Sun will evolve into a low-luminosity red giant similar in size to \id\ 
in roughly 7 billion years.

The Kepler planet search pipeline detected two planet 
candidates orbiting \id\ (designated as KOI-1241) \cite{borucki11} with 
periods of 10.50 and 21.41 days, a nearly
2:1 commensurability. The
observation of transit time variations caused by gravitational interactions
showed that the two candidates represent objects orbiting
the same star, and modeling of these variations led to
upper limits on their masses that place them
firmly in the planetary regime\cite{steffen12b}.
\id\ is the most evolved star observed by Kepler with more than one detected planet.

Transit observations lead to measurements of 
planet properties relative to stellar properties, and 
hence accurate knowledge of the host star is required to characterize the system. 
Asteroseismology enables inference of stellar properties through the measurement 
of oscillations excited by near-surface convection\cite{deridder09}. 
The power spectrum of the \id\ data after removing the planetary transits 
shows a regular series of peaks (Fig.\ 1), 
which are characteristic of stellar oscillations. 
By combining the measured oscillation frequencies with the effective temperature and chemical 
composition obtained from spectroscopy, we were able to
precisely determine the properties of the host star \cite{SOM}. 
\id\ is more than four times as large as the Sun and its mass is 30\% greater (Table 1).

Non-radial oscillations in evolved stars are mixed modes, behaving like 
pressure modes in the envelope and like
gravity modes in the core\cite{dziembowski01,montalban10}. 
Unlike pressure-dominated mixed modes, gravity-dominated mixed 
modes have frequencies that are shifted from the regular asymptotic spacing. Mixed modes 
are also approximately equally spaced in period \cite{beck11}.
We measured the average period spacing between dipole ($l=1$) modes 
in \id\ to be $50$~seconds, consistent with a
first ascent red giant\cite{bedding11}.

Individual mixed dipole modes are further split into multiplets as a result of
stellar rotation. Because the modes in each
multiplet are on average expected to be excited to very nearly equal amplitudes,
the observed relative amplitudes depend only on viewing angle relative
to the stellar rotation axis\cite{gizon03}.  For \id\, several mixed
dipole modes show triplets (Fig.\ 1).  A rotation axis perpendicular
to the line of sight (inclination $i=90^{\circ}$) would have produced
a frequency doublet, whereas a star viewed pole-on ($i=0^{\circ}$) would
have produced no visible splitting \cite{SOM}.  Therefore the observed
triplets are a clear signature of an intermediate inclination of the
stellar rotation axis with respect to the line of sight. This also
implies an intermediate inclination with respect to the planetary
orbital axes, which are known to be perpendicular to the line of sight
from the existence of transits.

For a quantitative measurement of the stellar spin-axis inclination, we modeled 
the six dipole modes with the highest signal-to-noise values. The fitted 
parameters included the frequency, height, width, rotational splitting and inclination 
for each mode. Three fitted modes (Fig.\ 1, middle panels) correspond to 
gravity-dominated mixed modes, whereas the other three multiplets (Fig.\ 1, bottom panels) 
are pressure-dominated mixed modes. The best fitting model
yields an inclination angle $i_{\rm g}=43\pm4^\circ$ for gravity-dominated modes
and $i_{\rm p}=51\pm4^\circ$ for pressure-dominated modes. 
Simulations confirmed that the inclination measurements 
are not strongly affected by the stochastic excitation of the oscillation modes, and
both inclinations are consistent with 
the coarser determination that can be made from estimates of the spectroscopic projected 
rotational velocity and the surface rotation rate \cite{SOM}. Furthermore, the measured 
splittings for gravity-dominated mixed modes are substantially higher than 
for pressure-dominated mixed modes, consistent with internal differential 
rotation in red-giant stars\cite{beck12}. Our observations thus reveal that \id\ rotates 
differentially with a rapidly spinning core, and that both the core and the envelope are 
(within 1.4-$\sigma$) mutually aligned and inclined by about 45$^\circ$ to the line 
of sight of the observer.

To measure the properties and orbital parameters of the planets, 
we used the stellar properties 
from asteroseismology to fit a model to the Kepler data that includes gravitational 
interactions between the planets, as revealed in the transit time variations 
(a ``photodynamical'' model; Fig.\ 2). In addition to the 
Kepler light curves, we 
obtained 10 high-precision radial velocity measurements using the 
HIgh-Resolution Echelle Spectrometer (HIRES) at the Keck 10-m 
telescope. The radial velocities show the Doppler signal 
of the transiting planets, as well as a slow velocity drift indicating a 
third, more massive companion in a wide orbit (Fig.\ 3). 
The combined fit of transit time variations and radial velocity data yields 
precise properties of the system (Table 1).
Both planets have densities consistent with gas-giant planets, 
and their radii are comparable to the radius of Jupiter $R_{\rm J}$ for planet c 
($R=0.88\pm0.04 R_{\rm J}$) and intermediate between Saturn and Neptune for 
planet b ($R=0.58\pm0.03 R_{\rm J}$). The planets are more than 30\%
larger than previously thought\cite{steffen12b}, 
because asteroseismology enables a more accurate measurement of the host star's
properties.

Further analysis also shows that the orbits of the planets are nearly
circular and co-planar. By itself the pattern of transit time
variations does not imply coplanar orbits, but in combination with the
radial velocity data the mutual inclination is required to be either
very low ($\lesssim 10^{\circ}$) or moderately high ($\gtrsim 60^{\circ}$) \cite{SOM}. 
We performed dynamical stability simulations using
initial conditions drawn from the posterior distribution of the
photodynamical model. The highly inclined solutions were
dynamically unstable on a timescale of $10^4$ years \cite{SOM}. Thus, 
the transiting planets in the Kepler-56 system are on co-planar
orbits that are misaligned with the equatorial plane of the host
star.

Several theories have been proposed to explain stellar spin-orbit misalignments.
Favored scenarios include dynamical 
perturbations such as Kozai cycles\cite{fabrycky07} and planet-planet 
scattering\cite{nagasawa08}. These scenarios would be consistent with the presence of a 
third companion, 
but would tend to randomize mutual inclinations of planets 
(and therefore lead to mutually inclined multi-planet systems) unless the perturbations 
occurred early enough for the inclinations to be damped by the protoplanetary disk.
Alternative tilting 
mechanisms invoke interactions of the stellar magnetic field with the proto-planetary 
disc\cite{lai11}, angular momentum transport within the star by internal gravity 
waves\cite{rogers12}, or tidal interactions in the early stages of star 
formation\cite{bate10}. These theories are consistent with a co-planar multi-planet 
system, but do not require the presence of a third companion on a wide orbit. 
Spin-orbit misalignments could also be produced through a scenario involving 
torques from nearby planets or 
companion stars in inclined orbits\cite{mardling10,batygin12}. 
Contrary to other scenarios, such a mechanism would naturally 
produce both a co-planar multi-planet system and a third companion in a wide orbit, 
as observed for Kepler-56.

The wide companion in the Kepler-56 system thus 
offers an intriguing explanation for the misalignment 
based on a scenario originally proposed for the transiting multi-planet system 
HAT-P-13\cite{mardling10}.
The radial velocity drift implies a third companion with the mass of a gas-giant planet within 
a few astronomical units, or a brown dwarf or star within several dozen astronomical units. 
In either case, if the third companion's orbit is itself inclined with
respect to the inner planetary orbits 
(for example through planet-planet scattering, if the companion is a planet), 
it could have torqued the  
orbits of the inner planets out of the equatorial plane of the host
star. The inner planetary orbits would stay aligned with one another because of 
strong coupling between their orbits, 
resulting in a misalignment of the two co-planar transiting planets with the host star.
Dynamical simulations that include a third companion in an eccentric 
orbit inclined to the equatorial plane of the host star confirm that 
such a mechanism can reproduce the architecture of the Kepler-56 system \cite{SOM}.

Obliquity measurements have long been considered as a powerful tool to
test planet formation theories\cite{queloz00,winn05}. 
In particular, observations of the Rossiter-McLaughlin effect have revealed that 
stars hosting hot Jupiters (mass $\gtrsim0.3$ times the mass of Jupiter, period $<10$ days) 
show a wide range of 
obliquities\cite{johnson09,winn10,triaud10,albrecht12}.
This finding has been interpreted as 
supporting evidence for dynamical perturbations as the origin of hot Jupiters, 
and against scenarios in which hot Jupiters migrate inward because of 
an interaction with the protoplanetary disk\cite{lin96}. 
This conclusion, however, relies on the assumption that
the stellar equator is a good tracer of the
initial orbital plane of the planet (and hence the protoplanetary disk), which 
has previously been called into question\cite{lebouquin09,watson11}. 
Important test cases are co-planar multi-planet systems which, 
if primordial alignments are common, should predominantly show low 
obliquities. Indeed, until now all transiting multi-planet 
systems have been found to be well-aligned\cite{hirano12b,sanchis12,chaplin12}. 

Although our observations do not constrain the 
primordial inclination of the protoplanetary disk of \id, 
they provide firm evidence 
that stellar spin-orbit misalignments are not solely confined to hot-Jupiter systems. 
Continued radial velocity measurements will reveal whether the third companion in the 
\id\ system is a planet (implying that the initial misalignment occurred after the 
planets formed) or a star (implying a primordial misalignment of the 
protoplanetary disk).

\vspace{0.7cm}
\noindent
\textbf{Acknowledgements:}
We gratefully acknowledge the entire Kepler team for making this paper possible. 
Funding for the Kepler Mission is provided by NASA's Science Mission Directorate. 
We thank E.\ Agol and D.\ Raggozine for helpful comments on the manuscript.
Supported by an appointment to the NASA Postdoctoral Program at Ames Research Center, 
administered by Oak Ridge Associated Universities through a contract with NASA (D.H.); 
a NSF Graduate Research Fellowship (K.M.D.);
a NSF Graduate Research Fellowship under grant DGE‐1144469 (B.T.M.);
the Netherlands Organisation for Scientific Research (S.H.);
BELSPO for contract PRODEX COROT (J.M.);
the NASA Kepler Participating Scientist program (R.S.-O., J.N.W.\ and E.B.F.);
NSF grant AST-1105930 (S.B.); 
and the David and Lucile Packard and Alfred P. Sloan foundations (J.A.J.);
J.A.C. is a Hubble Fellow of the Harvard-Smithsonian Center for Astrophysics.
Funding for the Stellar Astrophysics Centre is provided by  Danish 
National Research Foundation grant DNRF106. The research is supported 
by the ASTERISK (ASTERoseismic Investigations with SONG and 
Kepler) project funded by the European Research Council (grant agreement 267864). 

%
\begin{table}
\begin{center}
\vspace{0.1cm}
\begin{tabular}{l c}        
\textit{Host Star} & 	\\
\hline         
Radius ($R_{\odot}$)			& $4.23\pm0.15$		\\
Mass ($M_{\odot}$)				& $1.32\pm0.13$			\\
Mean Density (g cm$^{-3}$)			& $0.0246\pm0.0006$		\\
log [Surface gravity] (cgs)		& $3.31\pm0.01$		\\
Effective Temperature (K)		& $4840\pm97$		\\
Metallicity [M/H] (dex)			& $0.20\pm0.16$		\\
Age (Gyr)						& $3.5\pm1.3$	\\
Stellar Inclination (degrees)	& $47\pm6$		\\
\textit{Planet b}					&			\\
\hline
Time of Transit (BJD)		&	$2454978.2556^{+0.0056}_{-0.0057}$		\\
Orbital Period (days)		&	$10.5016^{+0.0011}_{-0.0010}$			\\
Semi-major axis (AU)		&	$0.1028^{+0.0037}_{-0.0037}$			\\
Radius (\rearth)			&	$6.51^{+0.29}_{-0.28}$					\\
Mass (\mearth)				&	$22.1^{+3.9}_{-3.6}$					\\
Mean Density (g cm$^{-3}$)	&	$0.442^{+0.080}_{-0.072}$				\\
\textit{Planet c}					&			\\
\hline
Time of Transit (BJD)		&	$2454978.6560^{+0.0057}_{-0.0055}$		\\
Orbital Period (days)		&	$21.40239^{+0.00059}_{-0.00062}$		\\
Semi-major axis (AU)		&	$0.1652^{+0.0059}_{-0.0059}$			\\
Radius (\rearth)			&	$9.80^{+0.46}_{-0.46}$					\\
Mass (\mearth)				&	$181^{+21}_{-19}$						\\
Mean Density (g cm$^{-3}$)	&	$1.06^{+0.14}_{-0.13}$					\\
\end{tabular}
\caption{\textbf{Properties of the \id\ system}. Host star properties were
derived using asteroseismology and high-resolution spectroscopy. The inclination angle 
was calculated as a weighted average of the inclination measured from gravity-dominated and 
pressure-dominated dipole modes, and includes uncertainties from
finite mode lifetimes \cite{SOM}. Because the orbits are not periodic, orbital 
periods and transit times for the planets 
refer to values at an arbitrary reference epoch [barycentric Julian Date (BJD) 2,454,970 BJD]. 
The mutual orbital inclination of the two planets is $5^{+3.4}_{-3.1}$ degrees at 
this epoch.}
\end{center}
\end{table}

\clearpage

\begin{figure}[h!]
\begin{center}
\resizebox{15.5cm}{!}{\includegraphics{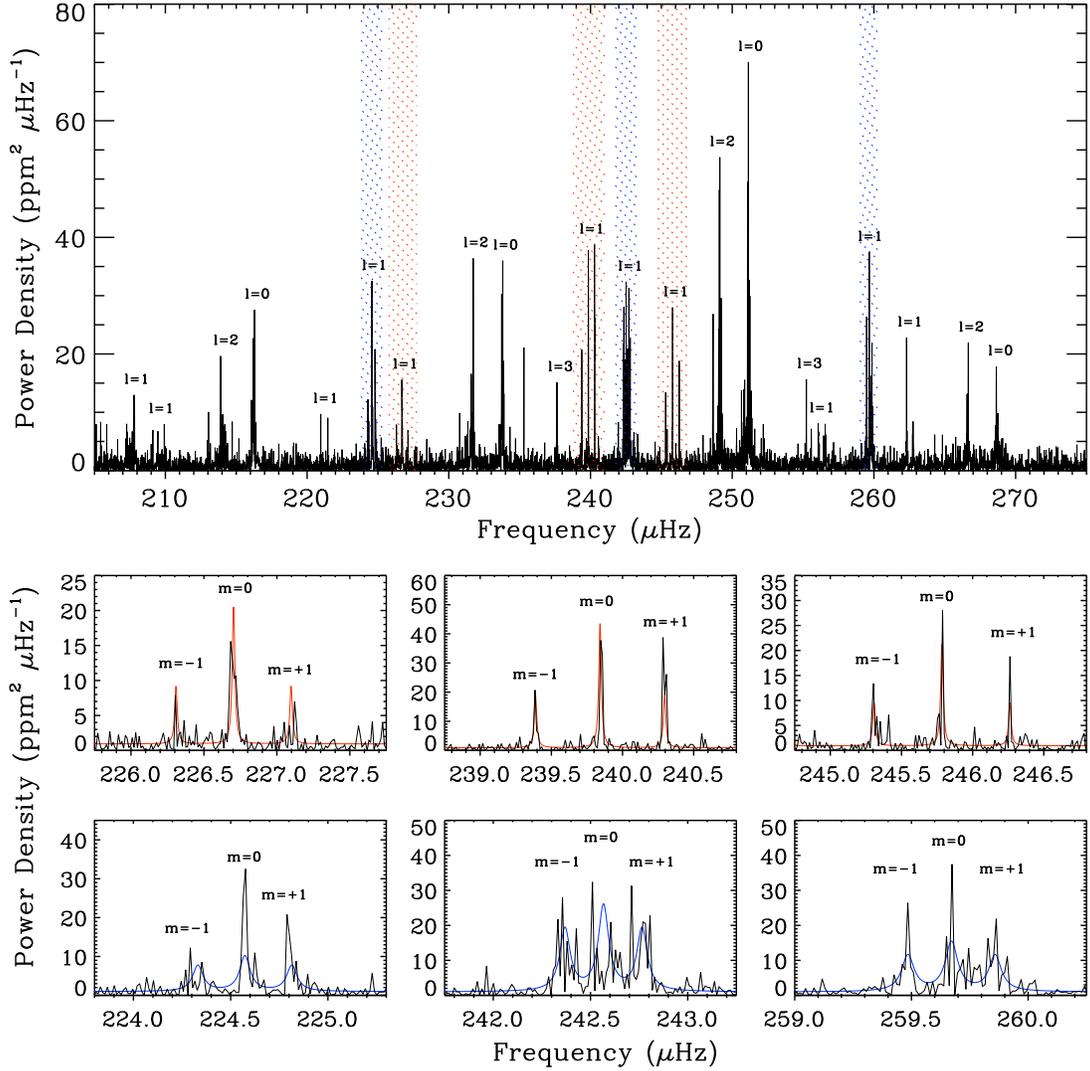}}
\caption{\textbf{Power spectrum analysis to measure the inclination of
    the stellar rotation axis.} \textit{Top panel:} Power spectrum
  centered on the frequency range with excited oscillations.  The
  spherical degree $l$ of each identified mode is
  indicated.  Red and blue areas highlight gravity-dominated and
  pressure-dominated mixed dipole modes, respectively. 
  \textit{Bottom panels:} Zoom on
  the mixed dipole modes highlighted in the top panel. Each mode is
  split into a triplet by rotation.  The azimuthal order $m$ of each
  component is indicated.  Red and blue lines show the modeled
  Lorentzian profiles.  The scatter in the data about the fitted
  model is due to the finite mode lifetimes \cite{SOM}.}
\end{center}
\end{figure}

\begin{figure}[h!]
\begin{center}
\resizebox{\hsize}{!}{\includegraphics{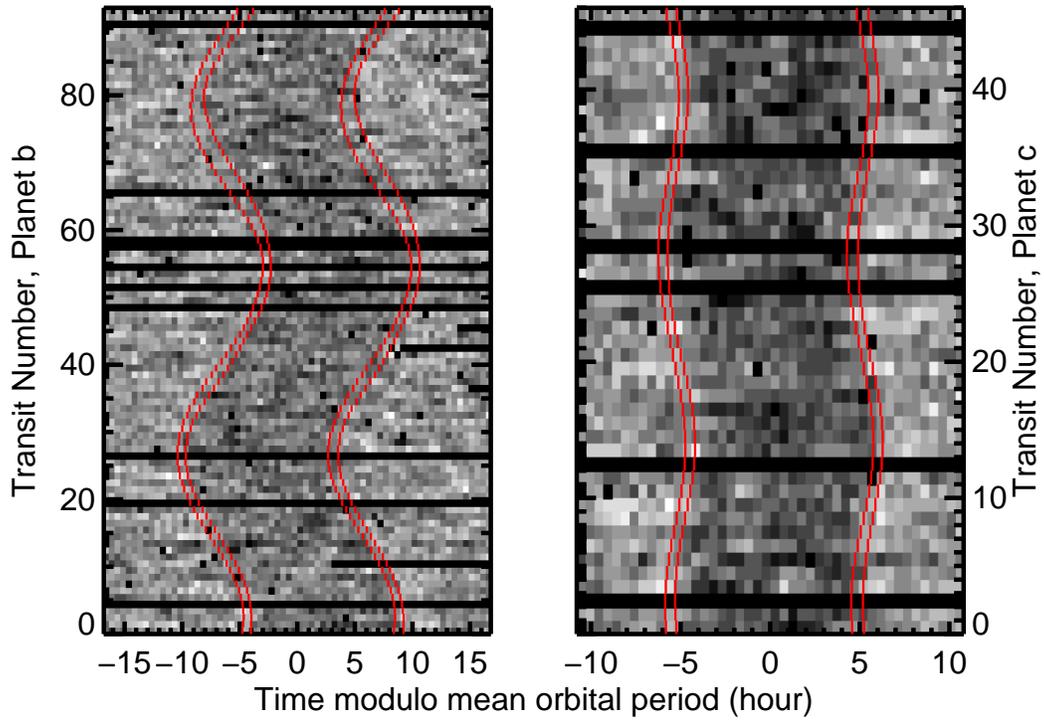}}
\caption{\textbf{Transit time variations of the inner
    planets}. Stellar intensity is plotted as a function of transit
  epoch and time modulo the mean orbital period near transits of 
  planet b (left) and c (right).  The red lines mark the 68\% confidence intervals
  for the start and end of each transit, according to the photodynamical
  model. }
\end{center}
\label{fig:river}
\end{figure}

\begin{figure}[h!]
\begin{center}
\resizebox{\hsize}{!}{\includegraphics{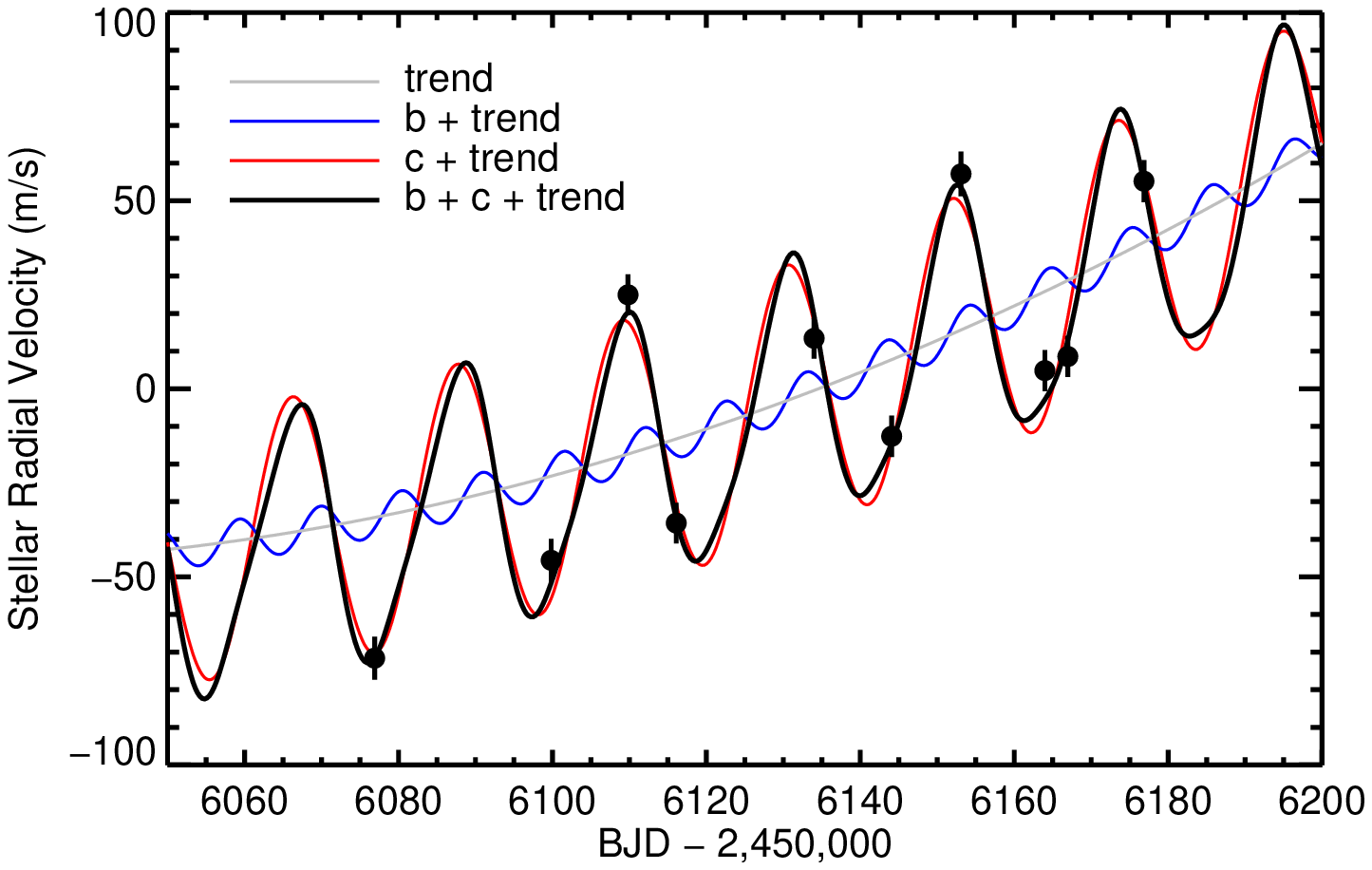}}
\caption{\textbf{Radial velocity variations.} Solid circles show the 
individual radial velocity measurements as a function of barycentric Julian date; 
the black solid line is the 
best-fitting photodynamical model to the combined Kepler and radial velocity data. 
Thin gray, blue and red lines show the individual components of the fit, 
which includes a radial velocity drift modeled as a quadratic function of time and 
radial velocity variations due to planet b and c.
The drift is attributed to a 
third, massive companion in a wide orbit.}
\end{center}
\end{figure}

\clearpage
\newpage
\setcounter{figure}{0}
\setcounter{table}{0}
\makeatletter 
\renewcommand{\thefigure}{S\@arabic\c@figure} 
\renewcommand{\thetable}{S\@arabic\c@table} 

\baselineskip12pt


\begin{flushleft} 
\Huge{\textbf{Supplementary Materials}}
\end{flushleft} 

\begin{normalsize}

\tableofcontents

\newpage

\section{Asteroseismic Data Analysis}

\subsection{Observations and Data Preparation}

The asteroseismic analysis is based on Kepler long-cadence data\cite{jenkins10} collected 
during Quarters 0--11, spanning a total of 977.8 days. We have used Kepler simple 
aperture photometry (SAP) for our analysis. Intensity differences between 
quarters were removed by fitting and correcting a 
linear regression to 10 day light curve segments 
before and after each quarterly gap. To correct remaining long-periodic 
instrumental trends, a quadratic Savitzky-Golay filter\cite{savitzky64} 
with a width of 2 days was applied. 

The sharp structure of transits in the time series can cause significant 
power leakage from low to high frequencies in the power spectrum, and hence they need to 
be corrected or removed prior to the asteroseismic analysis. Using the average 
ephemeris and orbital periods
identified by the Kepler planet search pipeline\cite{batalha12}, 
we removed data during transits from the time series. To account for 
transit timing variations, we used transit durations inflated to 19.5 hours for 
planet b and 16.3 hours for planet c to remove transits from the phase-folded light 
curve. Note that we have also repeated the analysis by discarding the transits according to 
the transit timings of the photodynamical model (\S\ref{sec:photodyn}), but 
found no significant difference in the results.

The removal of transits causes a reduction in duty cycle by $\sim$10\%.
Figure \ref{fig:win} shows the power spectrum of \id, with the spectral window 
after removing the transits overlaid on the oscillation mode with the
highest power. The spectral window (red) has sidelobes which are 
below 1\% in power, and hence make a negligible contribution to the power spectrum 
compared to the noise level. We conclude that the removal of transits has no 
significant impact on the results of the asteroseismic analysis. 

\begin{figure}[t!]
\begin{center}
\resizebox{13cm}{!}{\includegraphics{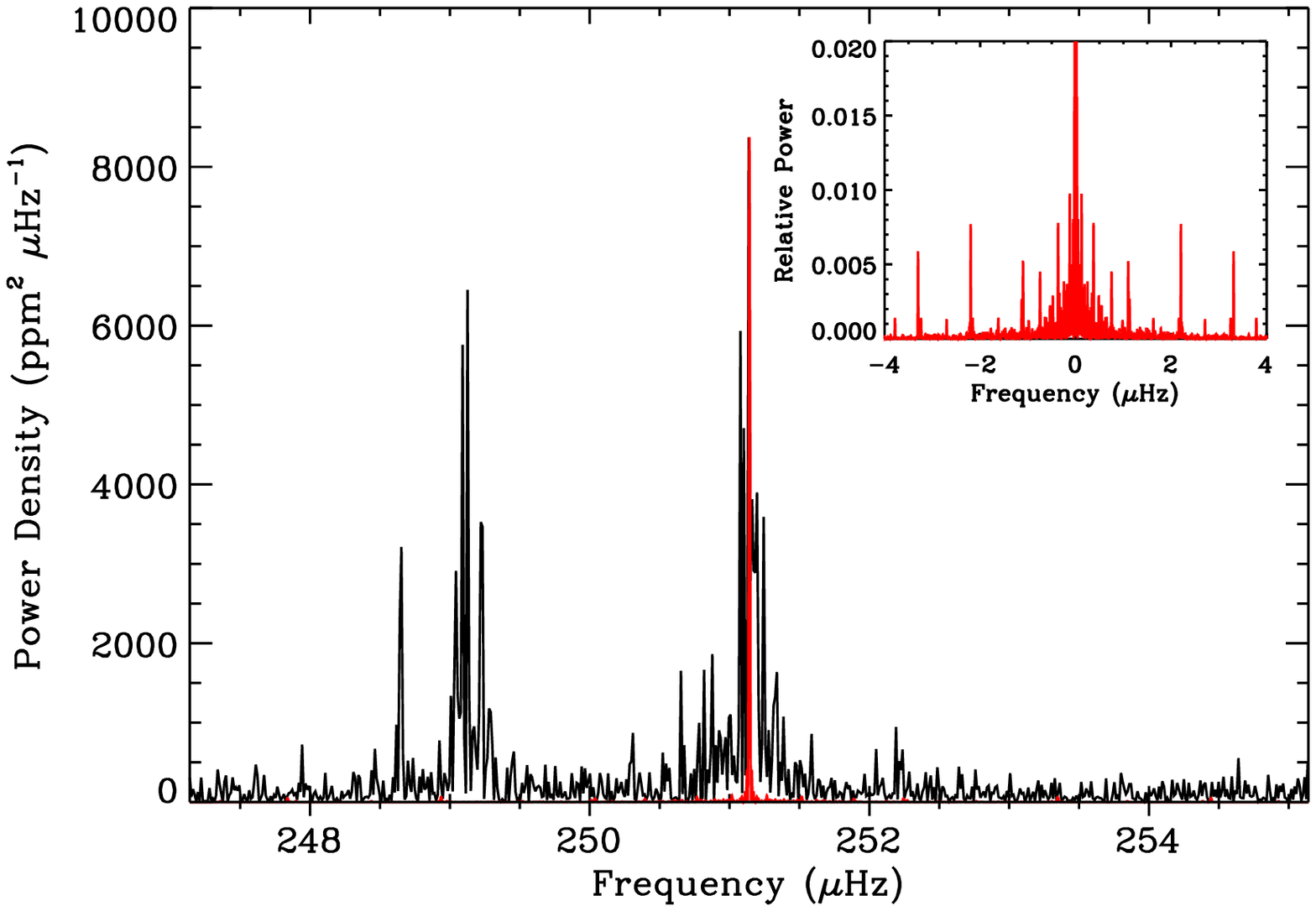}}
\caption{
A small part of the power spectrum of \id\ (black line), 
centered on the strongest oscillation mode. 
The red line shows the spectral window of the timeseries with transits removed,
scaled to the frequency and maximum value of the highest peak. 
The inset shows a close up of the spectral window, with the same $x$-axis range as the 
main panel and normalized to a height of 1. The sidelobes 
caused by the periodic transit removal are much lower than the overall noise level and hence 
negligible for the asteroseismic analysis.}
\label{fig:win}
\end{center}
\end{figure}

\subsection{Extraction of Oscillation Parameters and Frequencies}

To extract oscillation parameters characterizing the average properties of the power 
spectrum, we used automated analysis 
methods\cite{huber09,hekker10c} 
which have been thoroughly tested on Kepler data of other stars\cite{hekker11,verner11}. 
In brief, the power contribution due to granulation noise and stellar activity 
was modeled by a combination of power laws, and then corrected by dividing the power 
spectrum by the background model. Next, the frequency of maximum power 
(\numax) was measured by heavily smoothing the power spectrum or by fitting 
a Gaussian function to the power excess. Finally, the large frequency separation (\Dnu), 
i.e. the average separation of modes with the same spherical degree and consecutive radial 
order, was determined by computing an autocorrelation of the power spectrum or of the time 
series, and identifying the most significant peak. The 
high S/N of the \id\ data allowed a very precise determination of both quantities, 
yielding $\numax=244.3\pm1.4\,\muHz$ and $\Dnu=17.4\pm0.1\,\muHz$.

To extract individual oscillation frequencies we first
smoothed the background-corrected power spectrum with a Gaussian 
function with a FWHM of 1\muHz. The spectrum was then manually inspected for peaks 
significantly above the noise level (S/N $>$ 4), and for each identified mode 
a power-weighted centroid was calculated. To estimate uncertainties, 
Monte-Carlo simulations were performed by perturbing the power spectrum with random numbers 
drawn from a $\chi^2$ distribution with two degrees of freedom. 
For each iteration, the  
power weighted centroids were recalculated, and the uncertainty for each frequency was 
taken as the standard deviation of the resulting distribution. 
In an alternative approach, a 
statistical test was employed to identify frequencies with low probabilities 
of being due to noise \cite{hekker10b}. These frequencies were manually inspected and 
additional 
frequencies were selected if appropriate. The power spectrum was then fitted using a global 
fit, with the frequency, width and height as free parameters for each mode. The fitting was 
based on a maximum likelihood estimation and uncertainties were computed from the Hessian 
matrix. Finally, the uncertainties on the determined frequencies obtained using the methods 
described above were checked by 
performing a 
Markov-Chain Monte-Carlo analysis to fit a global model to the power spectrum 
\cite{handberg11}. 

\begin{table}
\begin{center}
\vspace{0.1cm}
\begin{tabular}{c c c c | c c c c}        
$f(\muHz)$  & $\sigma_{f} (\muHz)$ & $l$ & $m$ & $f(\muHz)$  & $\sigma_{f} (\muHz)$ & $l$ & $m$   \\
\hline         
   190.525  &      0.023  &       1  &      --  &    237.624  &      0.017  &	    3  &       0  \\
   192.402  &      0.029  &       1  &      --  &    239.380  &      0.007  &	    1  &      $-1$  \\
   196.888  &      0.019  &       2  &       0  &    239.848  &      0.007  &	    1  &       0  \\
   198.985  &      0.017  &       0  &       0  &    240.296  &      0.006  &	    1  &      $+1$  \\
   205.113  &      0.011  &       1  &      $-1$  &    242.363  &      0.011  &	    1  &      $-1$  \\
   205.437  &      0.009  &       1  &       0  &    242.566  &      0.015  &	    1  &       0  \\
   205.869  &      0.013  &       1  &      $+1$  &    242.749  &      0.013  &	    1  &      $+1$  \\
   207.730  &      0.025  &       1  &       0  &    245.326  &      0.011  &	    1  &      $-1$  \\
   209.055  &      0.018  &       1  &      $-1$  &    245.779  &      0.008  &	    1  &       0  \\
   209.463  &      0.011  &       1  &       0  &    246.278  &      0.012  &	    1  &      $+1$  \\
   209.996  &      0.018  &       1  &      $+1$  &    249.135  &      0.017  &	    2  &       0  \\
   214.017  &      0.025  &       2  &       0  &    251.150  &      0.016  &	    0  &       0  \\
   216.237  &      0.016  &       0  &       0  &    255.204  &      0.017  &	    3  &       0  \\
   220.965  &      0.011  &       1  &      $-1$  &    255.571  &      0.024  &	    1  &      $-1$  \\
   221.464  &      0.012  &       1  &       0  &    256.086  &      0.017  &	    1  &       0  \\
   221.926  &      0.011  &       1  &      $+1$  &    256.534  &      0.016  &	    1  &      $+1$  \\
   224.312  &      0.012  &       1  &      $-1$  &    259.486  &      0.009  &	    1  &      $-1$  \\
   224.582  &      0.007  &       1  &       0  &    259.687  &      0.011  &	    1  &       0  \\
   224.809  &      0.009  &       1  &      $+1$  &    259.842  &      0.011  &	    1  &      $+1$  \\
   226.340  &      0.013  &       1  &      $-1$  &    262.303  &      0.014  &	    1  &      --  \\
   226.699  &      0.007  &       1  &       0  &    262.746  &      0.015  &	    1  &      --  \\
   227.088  &      0.013  &       1  &      $+1$  &    266.617  &      0.019  &	    2  &       0  \\
   231.654  &      0.019  &       2  &       0  &    268.683  &      0.020  &	    0  &       0  \\
   233.760  &      0.015  &       0  &       0  &    277.538  &      0.024  &	    1  &       0  \\
\hline         
   213.035  &      0.022  &      --  &      --  &    235.307  &      0.020  &	   --  &      --  \\
   230.777  &      0.022  &      --  &      --  &    248.651  &      0.017  &	   --  &      --  \\
\hline         
\end{tabular}
\caption{Measured oscillation frequencies for \id. The spherical degree $l$ and 
azimuthal order $m$ is indicated for each frequency. Frequencies in the bottom two 
rows correspond to significant peaks for which no clear mode identification could 
be determined.}
\label{tab:freqs}
\end{center}
\end{table}

\begin{figure}[t!]
\begin{center}
\resizebox{\hsize}{!}{\includegraphics{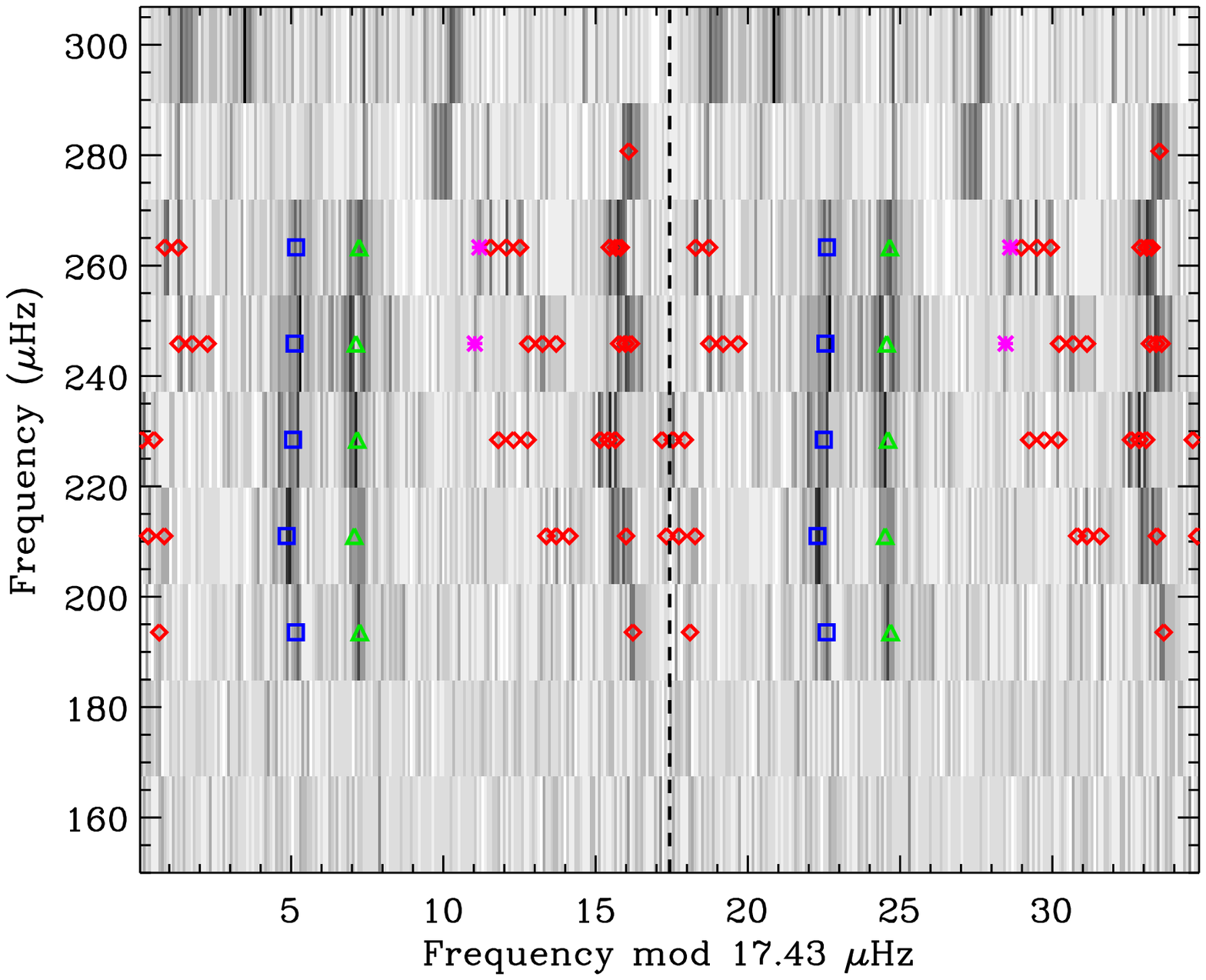}}
\caption{\'Echelle diagram of the background corrected power spectrum. Darker regions 
mark frequencies with higher power. Open symbols show 
extracted frequencies with spherical degrees $l=0$ (blue squares), $l=1$ (red diamonds), 
$l=2$ (green triangles) and $l=3$ (magenta asterisks). 
Note that for the extracted frequencies we have plotted the central frequency of each 
order on the vertical axis \cite{bedding10c}, and that the plot is duplicated 
past the vertical dashed line for clarity.}
\label{fig:echelle}
\end{center}
\end{figure}

Figure \ref{fig:echelle} shows an \'echelle 
diagram, which is calculated by plotting the power spectrum modulo the large 
frequency separation (hence stacking orders of equal spherical degree on top of each 
other). Oscillation modes with spherical degree $l=0, 1, 2$ and 3 are 
denoted by green triangles, blue squares, red diamonds and magenta asterisks.
The vertical ridge along which oscillation modes line up in the \'echelle diagram   
is much broader for $l=1$ modes than for $l=0$ and 2 modes, due to the presence of 
mixed modes \cite{bedding10b}.
Four orders contain $l=1$ modes with clear triplet structure due to rotational 
splitting \cite{beck12,deheuvels12,mosser12}, 
which allows for an identification of the azimuthal degree $m$. The extracted 
frequencies including a mode identification are listed in Table \ref{tab:freqs}.

\begin{figure}[t!]
\begin{center}
\resizebox{\hsize}{!}{\includegraphics{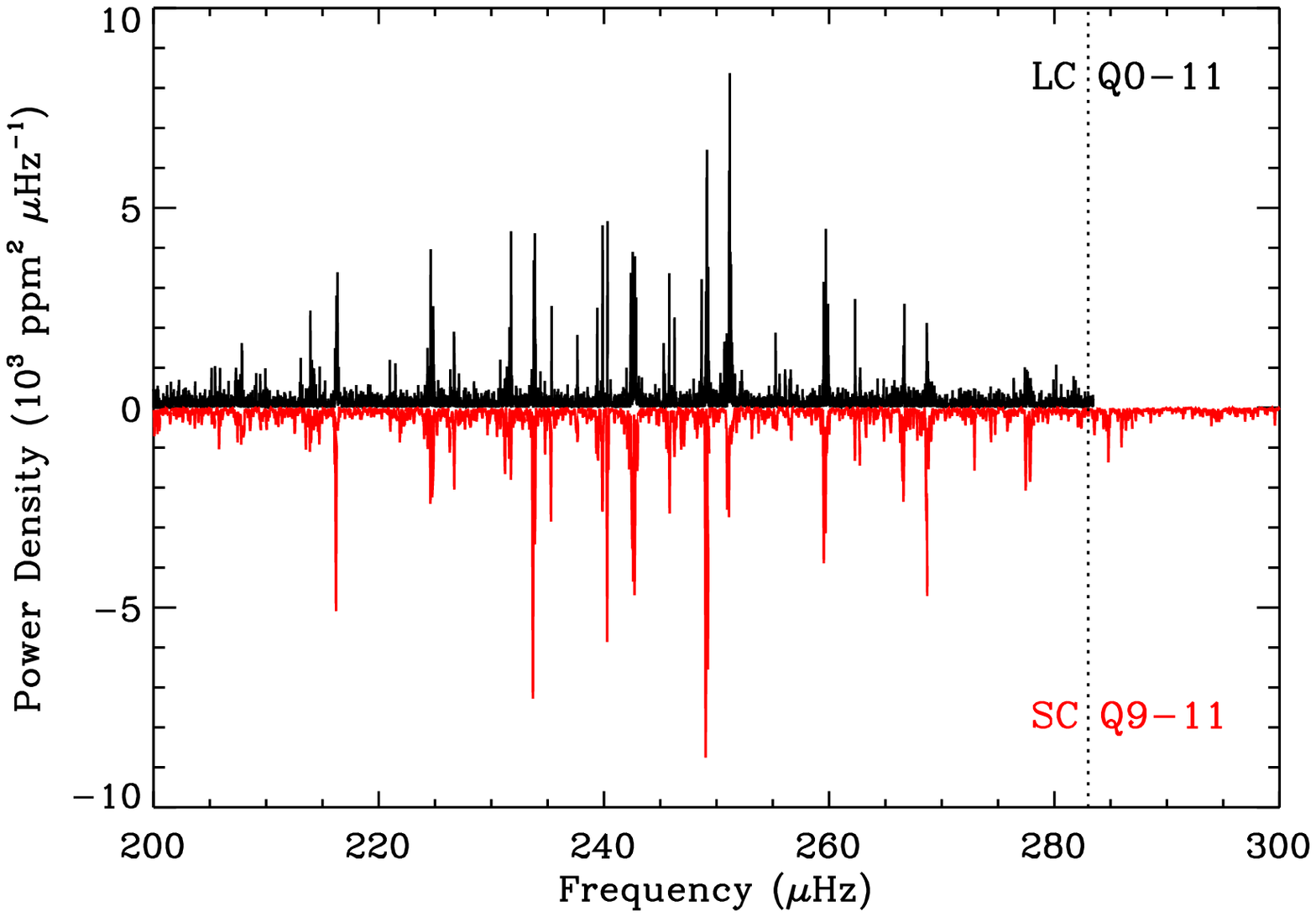}}
\caption{Comparison of power spectra calculated from Q0-11 long-cadence 
data (black) and Q9-11 short-cadence data (red). Note that the y-axis for the latter 
has been inverted for clarity. The vertical dotted line marks the long-cadence 
Nyquist limit. Both power spectra are nearly indistinguishable, showing that 
Nyquist effects are negligible when using long-cadence data for the asteroseismic 
analysis.}
\label{fig:comp}
\end{center}
\end{figure}

The proximity of the power excess to the long-cadence Nyquist limit 
(283\muHz) 
raises some concern about the effect of reflection of power at the Nyquist 
frequency on the extracted parameters, 
in particular for the determination of \numax. Although Q9-11 short-cadence data 
are available for \id, the four-times-higher frequency resolution in long-cadence data 
is essential for resolving the rotationally split multiplets. To test the influence of 
Nyquist effects,
Figure \ref{fig:comp} compares the power spectrum using 
long-cadence data to a power spectrum calculated using the Q9-11 short-cadence data. 
The comparison shows that there are no significant reflection effects in the long-cadence 
data and that, except for a few low-amplitude modes at the highest frequencies, all 
frequencies are well captured.
A re-determination of \numax\ and \Dnu\ using short-cadence data yielded nearly identical 
to those obtained using long-cadence data, but with higher uncertainties.
This confirms that the extracted oscillation parameters and individual 
frequencies using long-cadence data were not affected by Nyquist effects.

\subsection{Host Star Inclination}

\subsubsection{Power Spectrum Modeling}

The inclination of a rotating star can be determined by measuring the 
relative heights of rotationally split oscillation modes \cite{gizon03,chaplin12,gizon13}.
To measure the inclination of \id, we fit rotationally split Lorentzian 
profiles to the six strongest triplets of dipole modes in the power 
spectrum (see Figure 1). The 
model power $P$ as a function of frequency $\nu$ can be described as:

\begin{equation}
P(\nu) = \sum_{k=1}^{N} \sum_{m=-l}^{+l} \frac{ \epsilon_{lm}(i) h_{k} }{ 1+4(\nu-f_{k}-m s_{k})^2 \Gamma^{-2}} + n \: .
\end{equation}

Here, $h$ is the mode height, $f$ is the central mode frequency, 
$\Gamma$ is the mode linewidth, $s$ is the rotational splitting, and $n$ is an arbitrary 
noise floor in the power spectrum. 
The outer sum runs over the $N$ oscillation modes 
to be fitted, while the inner sum runs over the azimuthal order $m$ of each frequency.
The relative height of each 
component is given by $\epsilon_{lm}(i)$, which 
depends on the azimuthal order $m$, spherical degree $l$ and 
inclination angle $i$. 
For dipole modes ($l=1$), $\epsilon_{lm}(i)$ can be written as \cite{gizon03}:

\begin{equation}
\epsilon_{l=1,~m=0} = \cos^{2} i \: ,
\label{equ:eps1}
\end{equation}
\begin{equation}
\epsilon_{l=1,~m\pm1} = \frac{1}{2} \sin^{2} i \: .
\label{equ:eps2}
\end{equation}

Note that this formulation assumes that the intrinsic mode height is independent of $m$.
The amplitudes of the 
$m=\pm1$ components relative to the $m=0$ component then give a direct measure of the 
stellar inclination, independent of effects such as the Coriolis force or 
stellar limb darkening \cite{gizon03}. According to 
Equations (\ref{equ:eps1}) and (\ref{equ:eps2}) observations of frequency triplets 
can be used to immediately rule out edge-on or pole-on inclinations of the stellar spin 
axis.

We performed two fits, once using 
three gravity-dominated $l=1$ modes, and once using three pressure-dominated 
$l=1$ modes (see Figure 1). 
The power spectrum was first corrected for background contributions due to activity and 
granulation, 
as described in the previous section. For each mode, we fitted the central frequency $f$, 
rotational splitting $s$ and mode height $h$. The inclination $i$, linewidth $\Gamma$ and 
noise floor $n$ were 
assumed to be the equal for each set of three modes, yielding a total of 12 free 
parameters. The fit was performed using a Markov-Chain Monte-Carlo algorithm. 
The likelihood function $\mathcal{L}$ for fitting a power spectrum is given by 
$\chi^{2}$ statistics with 
two degrees of freedom and hence can be written as \cite{toutain94,benomar09b,gruberbauer09}:

\begin{equation}
\mathcal{L} = \prod_{\nu} \frac{1}{P_{\rm m}(\nu)} \exp\left(-\frac{P_{\rm o}(\nu)}{P_{\rm m}(\nu)}\right) \: .
\end{equation}

Here, $P_{\rm m}(\nu)$ is the power predicted by the model at a frequency $\nu$, and 
$P_{\rm o}(\nu)$ is the observed power. For the priors $p$ 
we assume Jeffreys priors for the mode heights:

\begin{equation}
{p(x)} = \frac{1}{x \ln({\frac{x_{\rm max}}{x_{\rm min}}})} \: ,
\end{equation}

\noindent
and uniform priors for the remaining parameters:

\begin{equation}
{p(x)} = \frac{1}{x_{\rm max}-x_{\rm min}} \: .
\end{equation}

Here, $x_{\rm min}$ and $x_{\rm max}$ are the minimum and maximum allowed values for a 
given parameter $x$.
We performed 
$5\times10^6$ iterations, and discarded the first 5\% of each chain. 
The best-fitting values and uncertainties were 
calculated as the median and 84.1 and 15.9 percentile of the marginalized posterior 
distribution for each parameter. Table \ref{tab:psfit} reports these values, and 
Figure \ref{fig:psposterior} shows the posterior distributions for 
each set of fitted gravity- and pressure-dominated $l=1$ modes. 
We have checked the results by sub-dividing the time series into two 
parts of equal length, and repeating the power spectrum analysis on these two 
independent datasets. The derived inclinations agreed well (within $6^{\circ}$ for 
p-dominated and g-dominated modes) with the results derived from the full dataset.

We note that the rotational splittings of gravity-dominated dipole modes are on average 
twice as large as the splittings measured from pressure-dominated dipole modes, consistent 
with other red giants observed by Kepler \cite{beck12,mosser12}. 

\begin{table}
\begin{center}
\vspace{0.1cm}
\begin{tabular}{l | c | c }        
Parameter  & g-dominated modes & p-dominated modes   \\
\hline   
$f_{1}(\muHz)$  & $226.7030^{+0.0066}_{-0.0069}$  & $224.5730^{+0.0077}_{-0.0076}$   \\
$f_{2}(\muHz)$  & $239.8421^{+0.0034}_{-0.0034}$  & $242.5669^{+0.0103}_{-0.0101}$  \\
$f_{3}(\muHz)$  & $245.7844^{+0.0040}_{-0.0038}$  & $259.6717^{+0.0086}_{-0.0084}$  \\
$s_{1}(\muHz)$  & $0.395^{+0.013}_{-0.018}$ 	  & $0.242^{+0.011}_{-0.011}$  \\
$s_{2}(\muHz)$  & $0.4534^{+0.0040}_{-0.0039}$    & $0.198^{+0.010}_{-0.010}$  \\
$s_{3}(\muHz)$  & $0.4773^{+0.0046}_{-0.0050}$    & $0.187^{+0.010}_{-0.010}$  \\
$h_{1}$  		& $36.0^{+20.6}_{-12.1}$  		  &   $22.3^{+6.6}_{-4.8}$\\
$h_{2}$  		& $78.3^{+41.4}_{-24.5}$  		  &   $59.9^{+20.3}_{-14.2}$\\
$h_{3}$  		& $37.5^{+19.0}_{-11.7}$ 		  &   $34.4^{+10.7}_{-7.8}$\\
$i$ (deg) 		& $42.5^{+4.4}_{-4.3}$			  &   $50.5^{+3.9}_{-4.0}$\\
$\Gamma(\muHz)$ & $0.0217^{+0.0056}_{-0.0048}$	  & $0.065^{+0.012}_{-0.011}$ \\
$n$				& $0.936^{+0.047}_{-0.045}$ 	  & $0.971^{+0.078}_{-0.073}$\\
\hline   
\end{tabular}
\caption{Results of fitting rotationally split Lorentzian profiles to two sets of three 
mixed $l=1$ modes. Fitted parameters are the central frequency $f$, the rotational 
splitting $s$, the mode height $h$, the inclination $i$, 
the linewidth $\Gamma$ and the noise floor $n$. Note that $h$ and $n$ are dimensionless 
quantities measured relative to the background.
The quoted values are the median as well as 84.1\% and 15.9\% 
confidence intervals.}
\label{tab:psfit}
\end{center}
\end{table}

\begin{figure}
\begin{center}
\resizebox{\hsize}{!}{\includegraphics{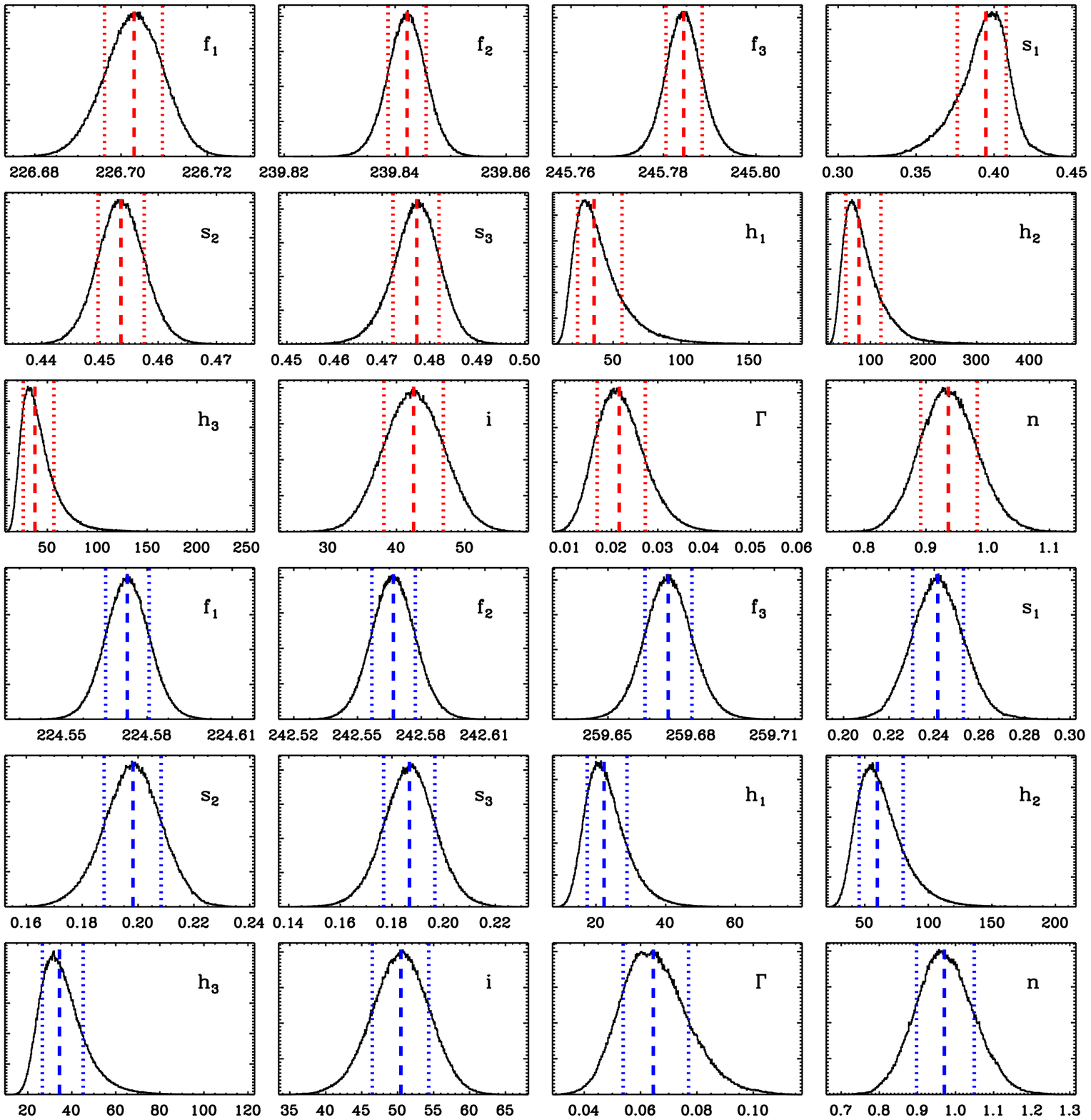}}
\caption{Posterior distributions of the MCMC analysis of
three gravity-dominated and three pressure-dominated rotationally-split dipole modes in the
\id\ power spectrum. Annotations in each panel follow the description of each parameter 
listed in Table \ref{tab:psfit}. Dashed lines show the median and dotted lines the 
84.1\% and 15.9\% confidence intervals, respectively. The three top rows show the posteriors
for  gravity-dominated modes (red lines) and the three bottom rows show the posteriors for 
pressure-dominated modes (blue lines).}
\label{fig:psposterior}
\end{center}
\end{figure}

\subsubsection{Finite Mode Lifetime Simulations}

It is important to examine how our measurement of the stellar inclination is affected 
by the finite lifetimes of the oscillation modes.
Solar-like oscillations are stochastically excited and damped \cite{houdek99,houdek06}, 
with mode lifetimes
ranging from a few days for main-sequence stars to 
several weeks or months for cool red giants 
\cite{chaplin09,dupret09,huber10,baudin11,appourchaux12,corsaro12}. 
The main effect of stochastic excitation 
is that solar-like oscillations are not described by a sinc function in the 
Fourier domain, but rather by a series of peaks modulated by a Lorentzian profile 
whose width depends on the lifetime of the modes. 
If the Lorentzian profile is not well resolved (i.e. the observation timebase is not 
much greater than the mode lifetime), the observed peaks will vary in height, 
depending on the time of observation.

The maximum mode lifetime for \id, based on the fit of Lorentzian
profiles, is about 170 days, 
indicating that the modes are not well-resolved by the observational timebase of 998 
days. To ensure that our measured inclination from the relative 
mode heights of rotationally split multiplets is not biased by finite mode lifetimes, 
we performed simulations as follows. Using the timestamps of the original \id\ 
observations, we generated synthetic timeseries by 
simulating a damped, harmonic oscillator with a given frequency, amplitude and 
mode lifetime \cite{chaplin97b}. The frequency and height of the simulated mode were set to 
typical values observed in \id\, and we added shot noise corresponding to the observed Kepler 
data. To simulate a rotationally split mode we 
added two additional modes spaced by an equal amount in frequency, 
and fixed the relative mode heights to the central mode for a given input inclination.
Figure \ref{fig:examples} shows several examples for synthetic power spectra 
over a range of mode lifetimes and inclinations for a single simulated mode in Kepler-56. 
A comparison with the underlying input models, shown in red, illustrates the effect of the 
finite mode lifetimes on the resulting spectrum.

We performed 2000 simulations by drawing random input values for inclinations and 
mode lifetimes from uniform distributions in the ranges 0--90 degrees and 30--300 days, and 
generating three rotationally split modes with typical frequencies and mode heights as 
observed in \id. For one half of the simulations we used a rotational splitting 
typical for pressure-dominated modes in \id, and for the other half we used 
rotational splittings typical for gravity-dominated modes.
For each simulation, we performed the same MCMC analysis as applied to the real data. 
Figure \ref{fig:simres} shows the determined inclinations compared to the input values, 
as well a histogram of the differences between output and input values. The results 
demonstrate that there is no bias introduced by finite mode lifetimes on the determination 
of the stellar inclination, and that inclinations are securely recovered for a wide range 
of input parameters. The residuals show a standard deviation of 5 degrees, 
in very good agreement with our estimated uncertainties for the original data. 
To account for finite mode lifetimes, 
we add in quadrature the scatter from our simulations to the uncertainty of the 
weighted average inclination from pressure- and gravity-dominated modes, 
yielding our final stellar inclination measurement for \id\ of $i=47\pm6^{\circ}$.

\subsubsection{Three-Dimensional Stellar Spin-Orbit Angle}

The three-dimensional angle $\psi$ between the stellar spin axis and the planetary orbital 
axes is given as \cite{fabrycky09}:

\begin{equation}
\cos \psi = \sin i \cos \lambda \sin i_{0} +  \cos i \cos i_{0} \: ,
\end{equation}

\noindent
where $\lambda$ is the sky-projected stellar spin-orbit angle, 
and $i_{0}$ is the angle between the line of sight and the orbital axis of the planet. 
The angle $\lambda$ can be measured through spectroscopic
observations of the Rossiter-McLaughlin effect or observations of planet - starspot 
crossings, but remains unconstrained in the asteroseismic analysis. 
If $i$ and $i_{0}$ are known, a lower limit of $\psi$ can be calculated:

\begin{equation}
\cos \psi < \sin i \sin i_{0} +  \cos i \cos i_{0} \: .
\end{equation}

Using the values for $i$ and $i_{0}$ derived from our asteroseismic and 
photo-dynamical analysis, we calculate $\psi > 37^{\circ}$ for \id.
For low eccentricity orbits, the lower limit on $\psi$ 
can be approximated as follows \cite{hirano12}:
 
 \begin{equation}
\cos \psi \lesssim \sin i + \frac{R_{\rm s}}{a_{\rm p}} \cos i \: .
\label{equ:psi}
\end{equation}

Here, $R_{\rm s}$ is the stellar radius and 
$a_{\rm p}$ is the semi-major axis of the planet.
Equation (\ref{equ:psi}) illustrates that
a large value of $i$ 
(i.e., the stellar rotation axis being nearly perpendicular to the line of sight) 
for a transiting system does not 
necessarily imply a stellar spin-orbit alignment, while 
a small value for $i$ always implies a 
stellar spin-orbit misalignment.

\begin{figure}
\begin{center}
\resizebox{\hsize}{!}{\includegraphics{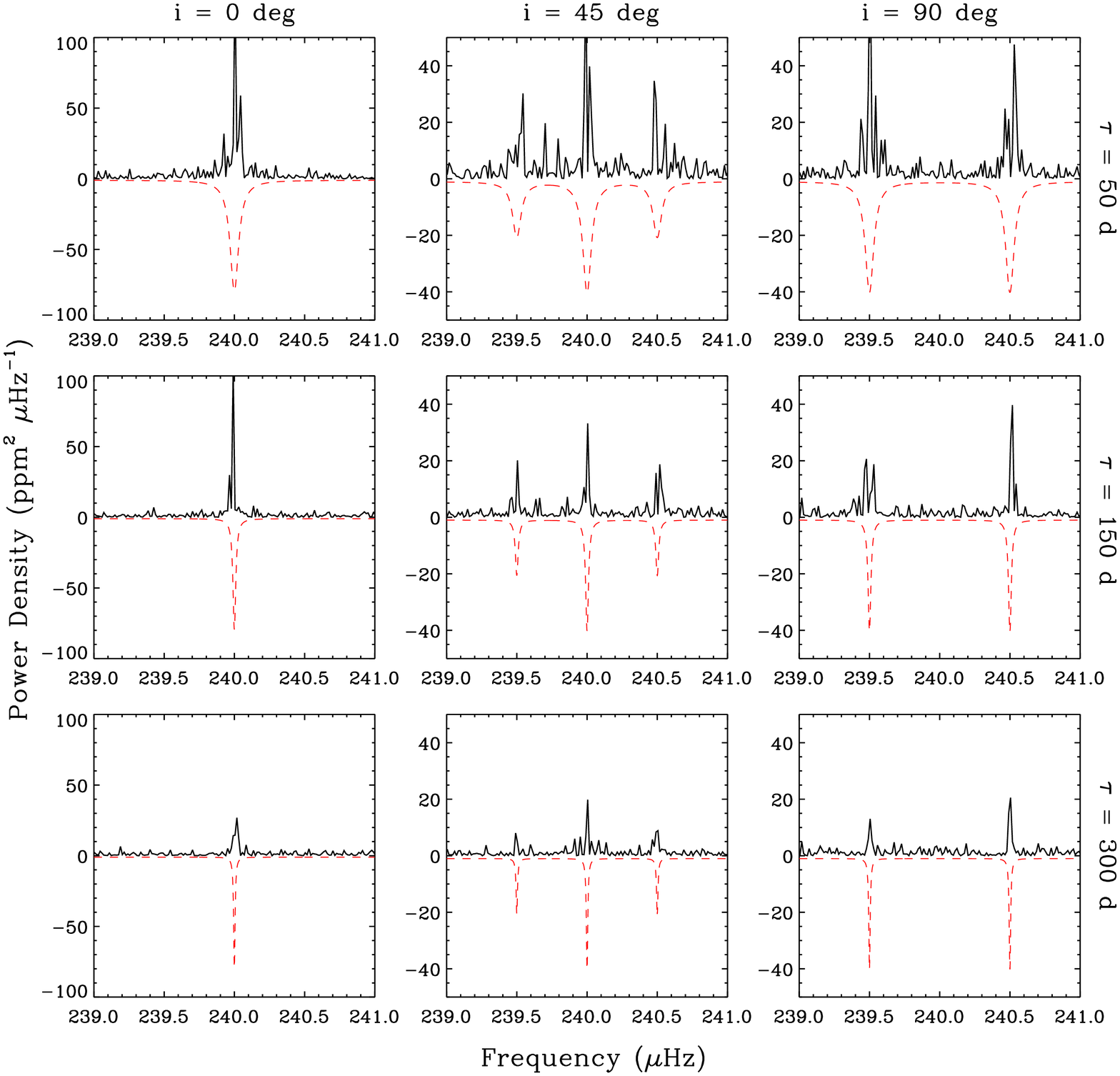}}
\caption{Simulations illustrating the influence of inclination and 
mode lifetime on observed power spectra. In each panel, red dashed lines show the input model and 
black lines show the calculated power spectrum of the simulated \id\ timeseries. 
Note that the input models have been inverted for clarity.
Mode frequencies, heights, and input 
shot noise were set to typical values for observations of \id.
The stochastic excitation of the 
modes causes the observed spectrum to scatter around the input spectrum. 
Note that only intermediate inclinations produce a distinct set of triplets, as 
observed for \id\ (see Equations (\ref{equ:eps1}) and (\ref{equ:eps2})).}
\label{fig:examples}
\end{center}
\end{figure}

\begin{figure}
\begin{center}
\resizebox{13cm}{!}{\includegraphics{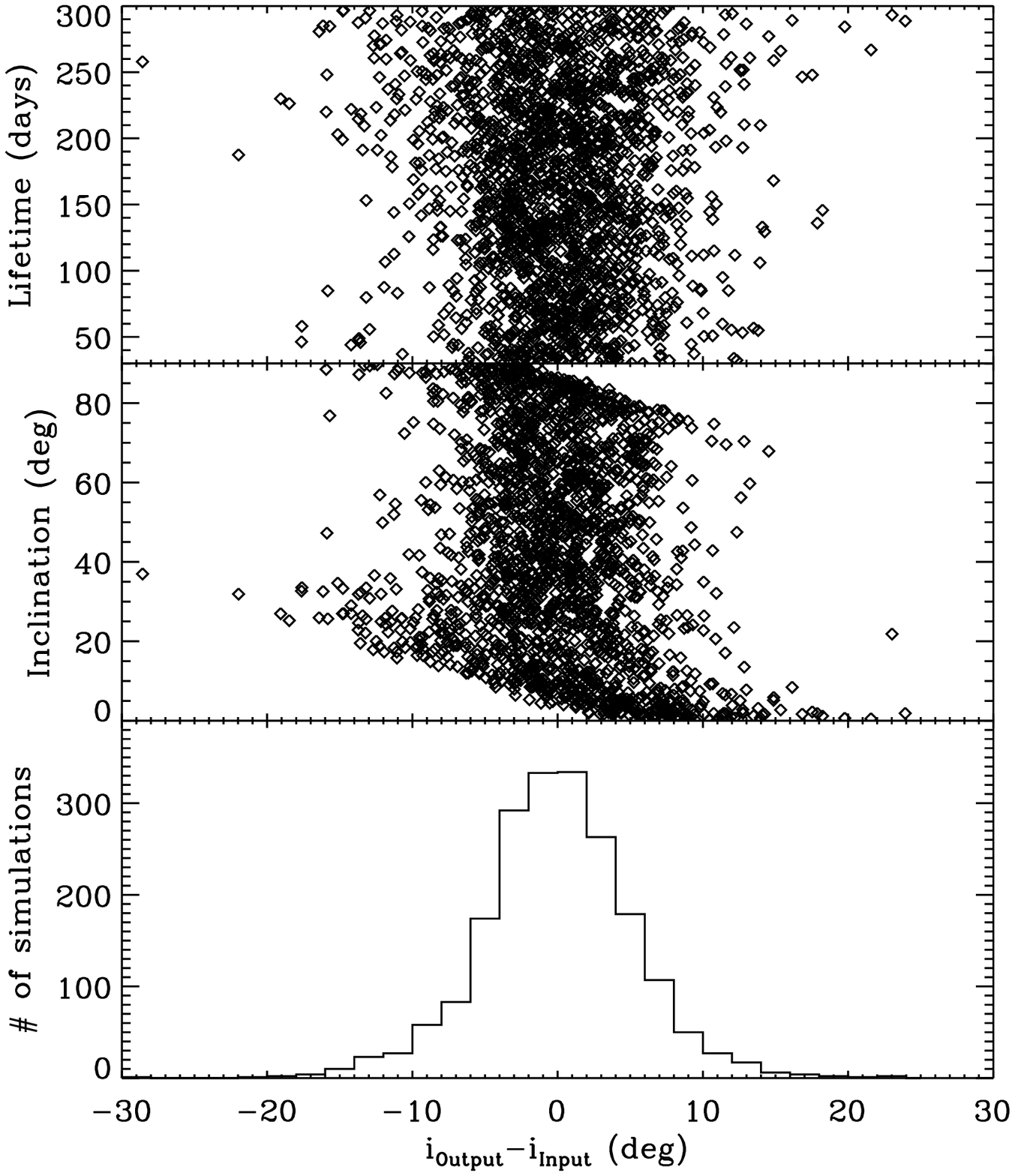}}
\caption{Simulation results to validate the measured inclination of \id. The top two panels 
show the input mode lifetime and input inclination versus the difference between the 
output and input inclination. 
Note that the sharp edges in the middle panel are caused by the condition that the 
output inclination is measured between 0 and 90 degrees.
The bottom panel shows a histogram of the 
differences between output and input inclination. The residual scatter over all 
mode lifetimes is $5^{\circ}$.}
\label{fig:simres}
\end{center}
\end{figure}

\section{Surface Rotation and Stellar Inclination from Starspots}
\label{sec:spotmodel}

Stars like Kepler-56 can have starspots that are carried across their surfaces by stellar 
rotation, which produces quasi-periodic flux variations. This variability was 
filtered out in order to study the stellar pulsations, which occur on much shorter 
timescales. However, rotational modulation due to spots can 
be used to measure the surface rotation rate and even to constrain  
the stellar inclination. Kepler simple aperture photometry (SAP) 
data cannot be used for this purpose because of artificial flux changes 
due to pointing drifts and other systematic 
effects \cite{jenkins10}. We therefore used the corrected flux series processed with the 
PDC-MAP algorithm 
\cite{stumpe12,smith12} to estimate the rotation period. This algorithm finds the 
systematic trends 
using a selected group of stars in each CCD module and uses that information to correct 
the light curves of all Kepler stars. The final product should mainly preserve
the astrophysical 
sources of variability. We applied a 3-sigma clipping algorithm to the PDC-MAP data with a 
12 hour-long moving-median filter, and also normalized each quarter by its median. The 
final flux series is shown in the upper panel of Figure \ref{fig:spots}, where 
the data have been binned 
to two points per day. Quasi-periodic variability of the order of $0.05-0.1$\% can be 
observed, with a period that does not seem to be strongly correlated with the quarter duration. 

We calculated a Lomb-Scargle periodogram and found a clear peak around 75 days 
(lower panel, Figure \ref{fig:spots}). This periodicity can be seen
also in the time series data. We interpret 
this periodicity as the signal introduced by spots rotating across the stellar disk. 
In the Lomb-Scargle periodogram we find the range of periods where the power is higher 
than half the peak 
power. This range is adopted as the 1-$\sigma$ uncertainties, and the center of the interval is 
the final value of the surface rotation period \cite{hirano12}. With this prescription, 
we find a rotation period of $74\pm3$ days, which corresponds to a frequency of 
$0.156 \pm 0.006 \muHz$. Combined with the rotational 
splitting of pressure and gravity dominated dipole modes, these observations show 
that the star rotates more slowly on the surface than within the
interior of the star \cite{beck12,deheuvels12,mosser12}. 
Using the radius of the star and the $v\sin i$ obtained 
from spectroscopy (\S\ref{sec:spectr}) we obtain an independent value of the stellar 
inclination of $i_{\rm s} = 36\pm25^{\circ}$, in agreement with the asteroseismic 
analysis. While the agreement is reassuring, the precision of the
$v\sin i$-based method is comparatively poor, and the accuracy of the
method is also questionable due to the difficulties of measuring
$v\sin i$ values as small as the one observed for Kepler-56 (see \S\ref{sec:spectr}).
Furthermore, the determined rotation period is close to the length 
of Kepler observing quarters and hence may be affected instrumental effects such 
as flux discontinuities between quarter boundaries.

\begin{figure}
\begin{center}
\resizebox{\hsize}{!}{\includegraphics{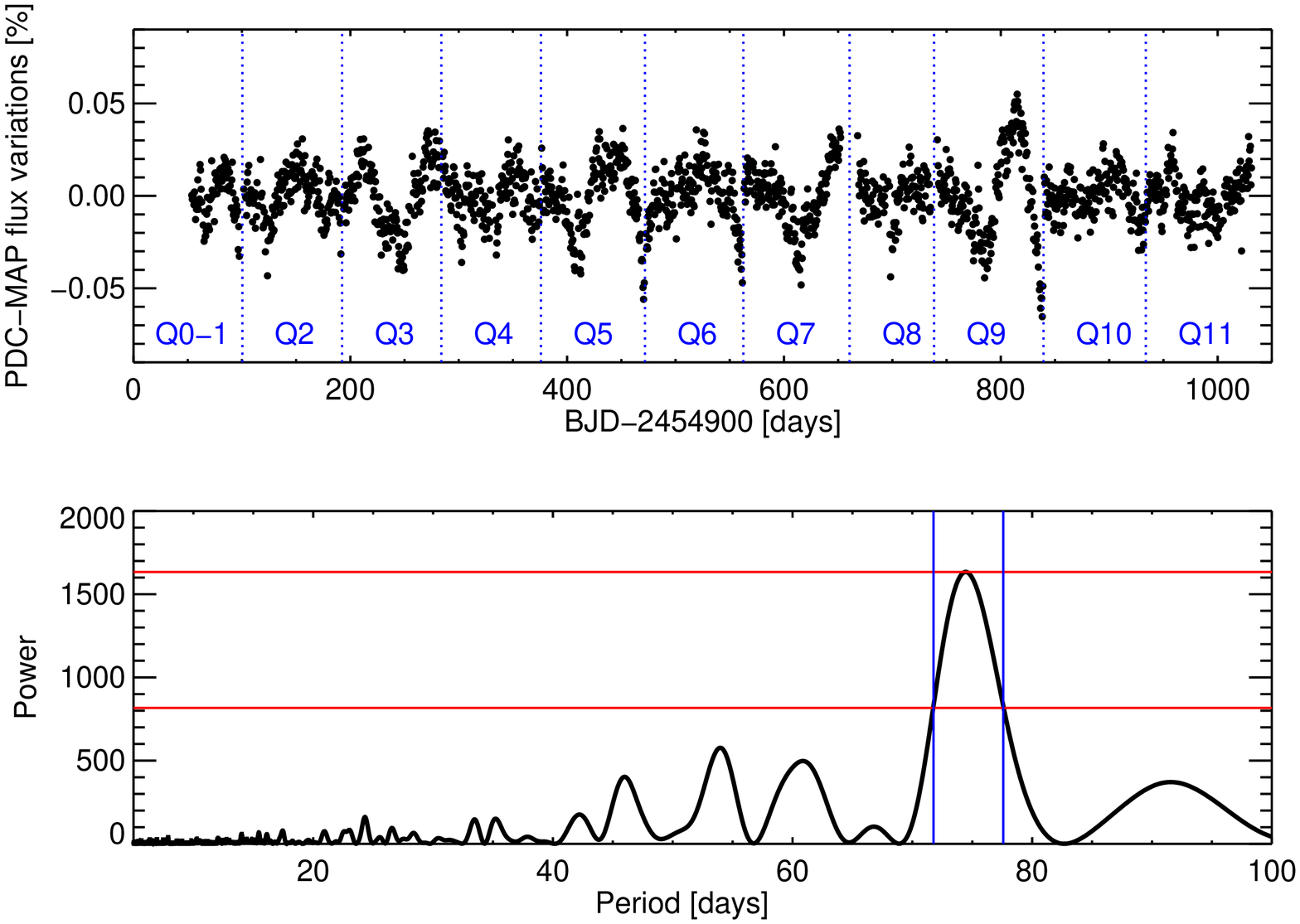}}
\caption{Quasi-periodic flux variations induced by starspots and rotation 
are used to obtain a surface rotation period and to check the stellar inclination.
Upper panel: The mean-normalized mean-subtracted PDC-MAP flux series is represented with 
black dots. The data are binned to show only two points per day. Each quarter is labeled 
in blue to show that the stellar variability is not strongly correlated with the 
quarter boundaries.
Lower panel: A Lomb-Scargle periodogram of the flux series shows a clear peak at 74 days, 
which we interpret as the surface rotation period. The red and blue lines mark the 
full-width at half-maximum of that peak.}
\label{fig:spots}
\end{center}
\end{figure}

\section{Host Star Properties}

\subsection{Atmospheric Parameters}
\label{sec:spectr}

We obtained spectroscopic follow-up observations of \id\ using the Fiber-fed \'{E}chelle 
Spectrograph (FIES) on the 2.5 m Nordic Optical Telescope (NOT) on La Palma, Spain 
\cite{djupvik10}. 
Three spectra were acquired in July 2011 with a resolution of R = 67,000 and an individual 
exposure time of 60 minutes yielding an average signal-to-noise ratio per resolution 
element of 47 in the MgB region. The stellar parameters were derived using the Stellar 
Parameter Classification pipeline (SPC) \cite{buchhave12}. In an initial analysis, effective 
temperature, surface gravity and metallicity were fit simultaneously to match the spectrum 
to a library of synthetic spectra. In a second iteration, the surface gravity was fixed 
to a value of $\logg = 3.29$, as determined from the asteroseismic gridmodeling analysis 
(see next section). This procedure was adopted to minimize potential correlations between 
Teff, log(g) and metallicity \cite{torres12}, and yielded final parameters of $\teff=4840\pm97$\,K, 
$\mh=+0.20\pm0.16$\,dex and $v\sin i=1.7\pm1.0$\,km\,s$^{-1}$. To account for systematic 
differences between different spectroscopic methods, the quoted uncertainties include 
contributions of 59\,K, 0.062\,dex and 0.85\,km\,s$^{-1}$ in \teff, \mh\ and $v\sin i$, 
respectively, which 
were added in quadrature to the formal uncertainties \cite{torres12}.

\subsection{Asteroseismic Grid-Modeling}

In a first step to estimate stellar properties using asteroseismology, we have used the 
asteroseismic observables \numax\ (the frequency of maximum power) 
and \Dnu\ (the average separation between modes of the same spherical degree and 
consecutive radial order). It can be shown that \numax\ and \Dnu\ are approximately related 
to stellar properties as follows \cite{ulrich86,brown91,kb95}:

\begin{equation}
\Delta\nu \approx \frac{({M/M_{\sun}})^{1/2}}{(R/R_{\sun})^{3/2}} \Delta\nu_{\sun} \: ,
\label{equ:dnu}
\end{equation}

\begin{equation}
\nu_{\rm max} \approx \frac{ M/M_{\sun}}{(R/R_{\sun})^{2}\sqrt{T_{\rm eff}/T_{\rm eff,\sun}}} \nu_{\rm max,\sun} \: .
\label{equ:nmax}
\end{equation}

Given an estimate of $T_{\rm eff}$, Equations (\ref{equ:dnu}) and (\ref{equ:nmax}) can 
be solved to obtain radius and mass, in the so-called direct 
method \cite{kallinger10c,mosser10}. 
Alternatively, \Dnu\ and \numax\ can be used in combination with 
evolutionary tracks, spectroscopic temperatures and metallicities to estimate 
stellar properties \cite{stello09,gai11,huber13}. 
Equations (\ref{equ:dnu}) and (\ref{equ:nmax}) 
have been tested observationally using 
eclipsing binary systems, Hipparcos parallaxes, and 
long-baseline interferometry \cite{miglio11,huber12b,silva12}, and have also been 
supported theoretically \cite{stello09,belkacem11}.
For evolved stars, Equations (\ref{equ:dnu}) and (\ref{equ:nmax}) have generally been 
found to yield radii and masses accurate to 5\% and 10\%, respectively 
\cite{belkacem12,miglio13}.

To derive fundamental properties, we have employed the grid-based method 
to match the spectroscopic and asteroseismic parameters to 
a variety of evolutionary tracks \cite{dasilva06,stello09,basu11,miglio12,silva13}.
The solar reference parameters used were $\numax=3090\,\muHz$ and $\Dnu=135.1\,\muHz$ 
\cite{huber11b}.
We note that the recently proposed revision of Equation (\ref{equ:dnu}) 
\cite{mosser13} has negligible influence on our results, and that the stellar properties 
derived from scaling relations are only used as an input for a more detailed analysis 
using individual oscillation frequencies (see next section) and 
to reduce degeneracies in the spectroscopic analysis (see previous section).

\subsection{Individual Frequency Modeling}

We have used the ATON code \cite{ventura08} to compute a grid of 
stellar interior models with masses in the range $1.26-1.56$\,\msun\ in steps of 
$0.02\,\msun$, 
helium mass fractions of $Y=0.27-0.33$ in steps of 0.01, metal mass fractions of 
$Z=0.028-0.030$ in steps of 0.001 and mixing length parameters $\alpha_{\rm MLT}=1.9, 2.05$ 
and $2.2$. For each track we computed adiabatic oscillation frequencies using 
LOSC \cite{scuflaire08,montalban10} for all models having a 
large frequency separation within 10\% of the observed value.

Model frequencies were corrected for near-surface effects 
\cite{kjeldsen08}. The power-law correction was applied to both 
radial and non-radial frequencies. Since the latter may have inertias considerably 
different to those of radial modes, the surface correction for non-radial modes was 
multiplied by $Q_{n,l}^{-1}$, where $Q_{nl}$ is the ratio of the mode inertia of the mode to 
that of the closest radial mode \cite{aerts08}.
To explore uncertainties in the exponent $b$ describing the power-law correction we 
considered $b=3,6$, and $8$.
To match the model frequencies to the observed frequencies we evaluated 
the reduced $\chi^{2}$ for the frequencies and the spectroscopic constraints 
separately. For the best matching models, the contribution of the 
spectroscopic constraints to the reduced $\chi^{2}$ is typically lower than 1.

Figure \ref{fig:modelechelle} shows an \'echelle diagram of the best-fit model compared 
to the observed 
frequencies. The match of both radial and non-radial frequencies to the observations 
is very good, and 
in particular reproduces the mixed dipole modes. We note that our best-fit 
models indicate that the trapping between the pressure-mode and gravity-mode cavities is 
strong enough for some $l=2$ mixed modes to 
have relatively low inertias, and therefore possibly excited to observable amplitudes.
Some of the additional modes with no clear 
identification (see Table \ref{tab:freqs}) may be compatible with mixed $l=2$ modes.

The properties of the best-fitting model are 
$M=1.32\pm0.13\,M_{\odot}$, $R=4.23\pm0.15\,R_{\odot}$ and
$\rho = 0.0246\pm0.0006$\,g\,cm$^{-3}$, 
with an age of $3.5\pm1.3$\,Gyr. 
The helium mass fraction, iron mass fraction and mixing length parameter of the best-fitting 
model are $Y=0.29$, $Z=0.03$, and $\alpha_{\rm MLT}=2.2$.
Fully consistent results were derived with an independent analysis 
using ASTEC models \cite{CD08a,CD08b,CD10}.
Uncertainties on the properties were estimated by adopting the 
fractional uncertainties of the grid-based method described in the previous section. 
These estimates encompass the properties of best-fit models obtained by fitting 
individual frequencies and making different assumptions on the surface-correction 
term (see above). 
Importantly, we note that the mean stellar density derived using individual model 
frequencies is 5\% higher than the density derived using Equation (\ref{equ:dnu})
($\rho=0.0234\pm0.0003$\,g\,cm$^{-3}$). 
This result has been confirmed using other techniques to model individual frequencies, 
and is in-line with previous studies showing deviations of Equation (\ref{equ:dnu}) 
from models for evolved stars \cite{white11,miglio13}.

\begin{figure}
\begin{center}
\resizebox{\hsize}{!}{\includegraphics{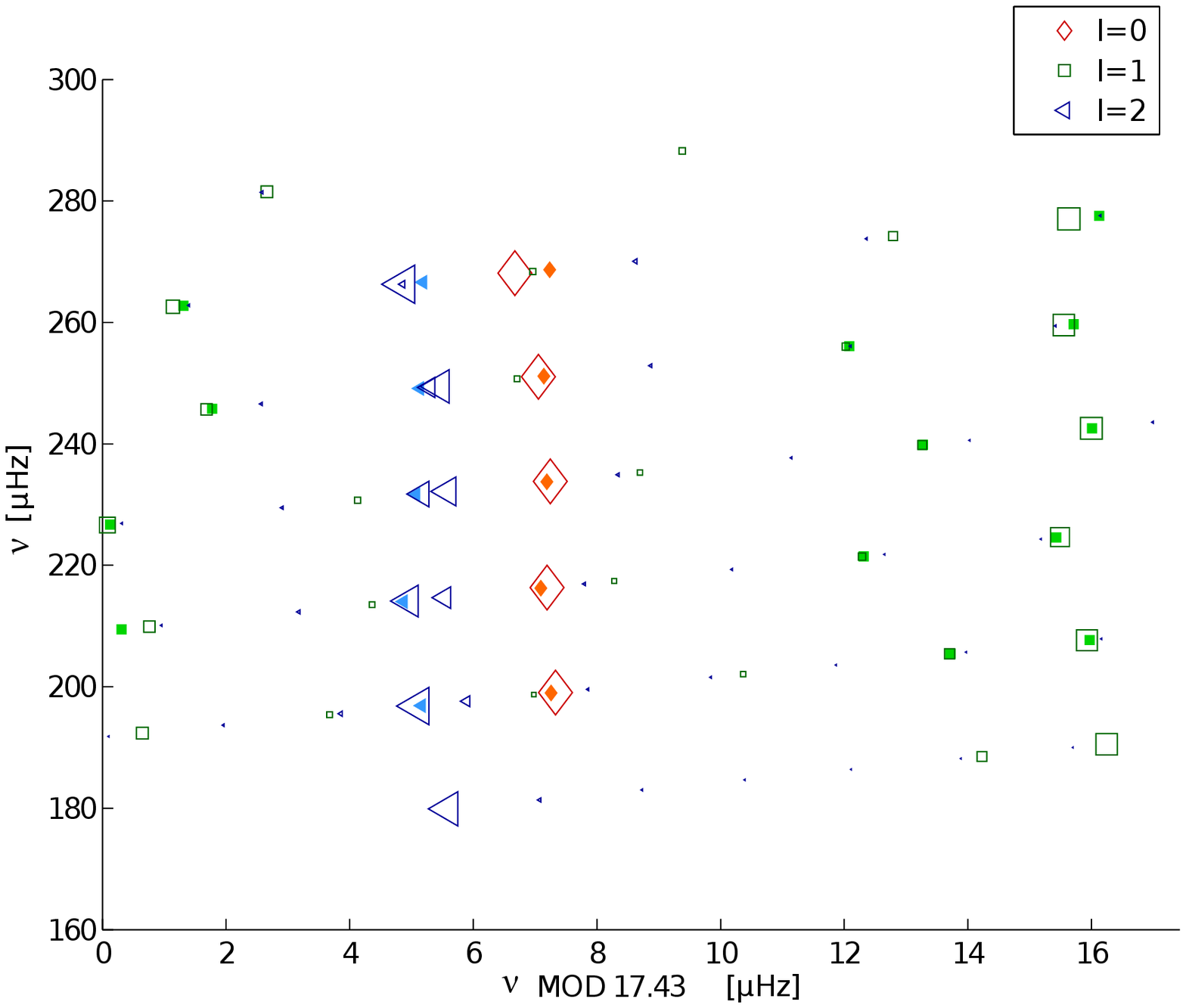}}
\caption{\'Echelle diagram comparing the observed frequencies (filled symbols) with 
theoretical frequencies of the best-fitting model (open symbols). 
Modes of different spherical degree are shown as diamonds ($l=0$), 
squares ($l=1$) and triangles ($l=2$). 
The size of the open symbols is inversely proportional to $E^{1/2}$, where $E$ is the mode 
inertia \cite{CD04}.
Note that 
rotationally split components ($m \neq 0$) are not included in the merit function 
and hence are not plotted in the diagram.}
\label{fig:modelechelle}
\end{center}
\end{figure}

To illustrate the evolutionary state of \id, 
Figure \ref{fig:hrd} shows evolutionary tracks 
from the BaSTI database \cite{basti} for the measured metallicity of \id, quadratically 
interpolated 
to a fine grid in stellar mass. The red box shows the position of \id\ as determined 
from the asteroseismic analysis of individual frequencies and spectroscopic follow-up. 
Additionally, green and blue models highlight the 1-$\sigma$ constraints from \Dnu\ and 
\numax, as used in the previous section. 

\begin{figure}
\begin{center}
\resizebox{\hsize}{!}{\includegraphics{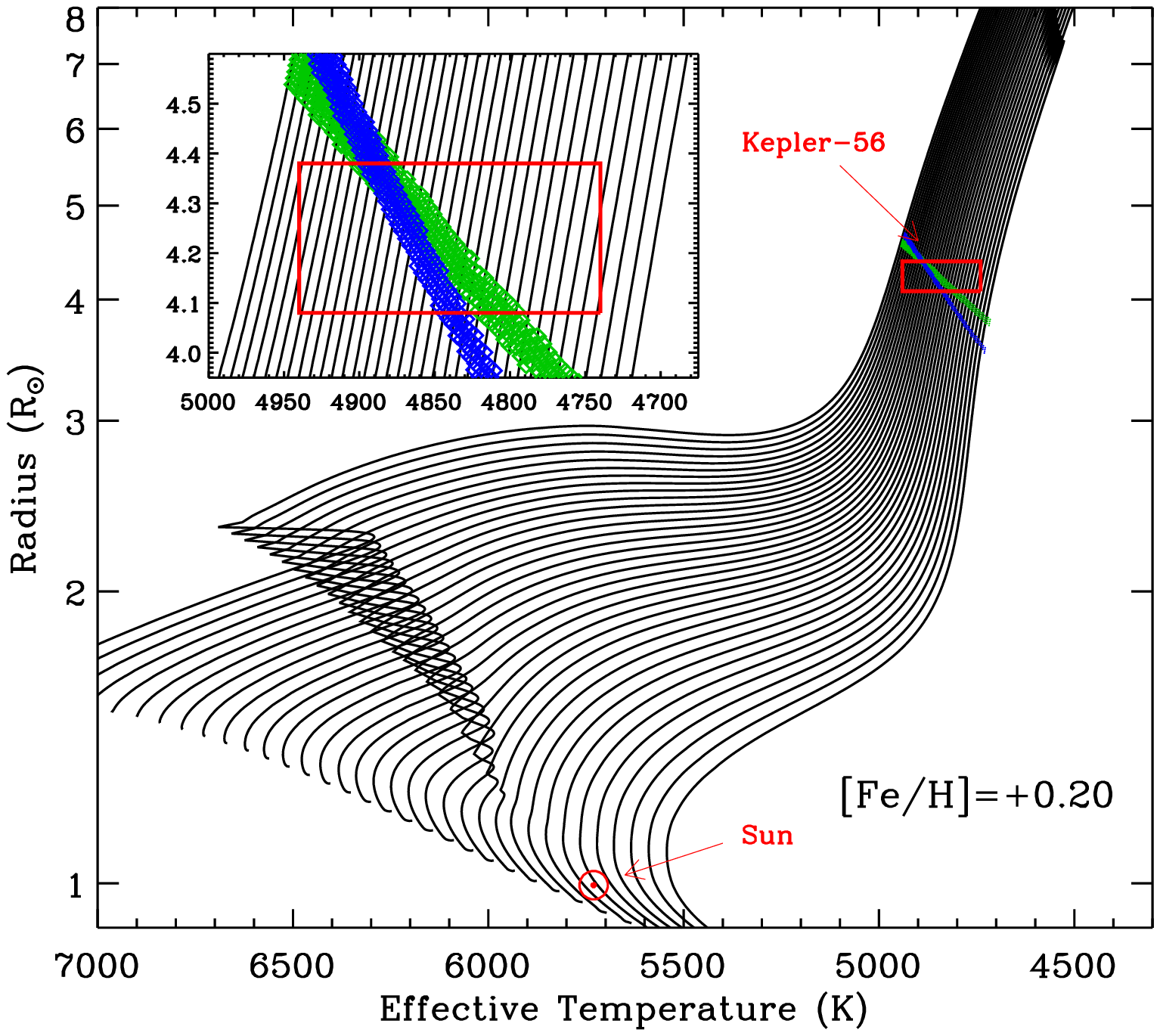}}
\caption{BaSTI evolutionary tracks for a metallicity of [Fe/H] = 0.20 and masses 
ranging from 0.9 to 1.6 solar masses with a stepsize of 0.02. Models fullfilling 
1-$\sigma$ observational constraints for the 
large frequency separation \Dnu\ (green) and the frequency of maximum power \numax\ 
(blue) are highlighted in the plot. Red lines show the 1-$\sigma$ error box in radius and 
temperature  
derived from asteroseismic modeling of individual frequencies and the spectroscopic 
analysis with asteroseismic constraints.
The position of the Sun is shown in the 
bottom part of the plot. The inset shows a close-up of the position of \id.}
\label{fig:hrd}
\end{center}
\end{figure}

\section{Radial Velocity Data}

We obtained spectroscopic observations of Kepler-56 at Keck Observatory (Mauna Kea, Hawaii) 
using the HIgh-Resolution Echelle Spectrometer (HIRES) \cite{vogt94} with the standard 
observational  setup used by the California Planet Survey \cite{howard10}. All 
observations were made
with an iodine cell mounted directly in front of the spectrometer entrance slit. The iodine
absorption lines observed with the stellar spectrum provide a precise wavelength scale to
measure Doppler shifts and place constraints on the shape of the HIRES instrumental profile 
at each observing epoch \cite{marcy92}.

Because of the star's relative faintness ($m_{\rm V} = 12.8$) we used the C2 decker, corresponding to a
sky-projected size of $14\farcs 0$ by $0\farcs 851$. The increased height of the C2 decker, compared to 
the shorter B5 decker normally used for brighter stars, allows for sky subtraction and 
provides a resolving power of $R = \lambda / \Delta \lambda \approx$~55,000. We obtained 
a total of 10
observations with exposure times ranging from 750 to 1800 seconds, resulting in 
signal-to-noise ratios between 50 and 90 at 550\,nm.

In each observation, we determined the radial velocity of Kepler-56 relative to a 
``template'' observation of the star with its instrumental profile removed through 
deconvolution. Two templates were collected of the star, each without the presence 
of iodine in the light path. The same decker was used for the templates 
as for the other observations.
The radial velocity measurements, times of observation, 
and internal uncertainties are listed in Table \ref{tab:rv}.

\begin{table}
\begin{center}
\vspace{0.1cm}
\begin{tabular}{l c c c}        
JD $-$ 244000  & RV (m~s$^{-1}$) & Unc.~(m~s$^{-1}$) & SNR   \\
\hline   
16076.904083 & 	-71.6087 & 	2.4468 & 		62 \\
16099.840740 & 	-45.5976 &	2.4815 &		62	\\
16109.824833 & 	24.9774  &	1.7344 &		88	\\
16116.089211 &	-35.6964 &	1.5950 &		89	\\
16133.999602 &	13.4154  &	1.6138 &		89	\\
16144.079281 & 	-12.6180 & 	1.9468 &		84	\\
16153.086751 &	57.1263  &	2.9634 &		52	\\
16163.980793 &	4.8418 	 &  1.8606 &		89	\\	
16166.962497 &	8.5878 	 &  1.7946 &		89	\\
16176.855891 &	55.1829  &	2.1804 &		65	\\
\hline   
\end{tabular}
\caption{Radial velocities for Kepler-56. The uncertainties reported
  in the third column are formal
measurement uncertainties and do not include the effects of stellar chromospheric ``jitter'' 
on our observations.}
\label{tab:rv}
\end{center}
\end{table}

\section{Photodynamical Modeling}
\label{sec:photodyn}

The times of transit of the two planets are not strictly periodic owing to planet-planet 
dynamical interactions.  These deviations may be interpreted to infer bulk and orbital 
properties including, for example, a combination of planetary mass and orbital 
eccentricity, or, the mutual inclination between the planetary orbits 
(the stellar density from asteroseismology also helps to constrain the vectorial eccentricity 
component $e\sin\omega$ for each orbit, where $e$ is the eccentricity and 
$\omega$ the argument of periastron).  However, the transit times are difficult to 
estimate at individual epochs owing to correlated noise in excess of the photon noise.  
To attempt to resolve this, we fit all transit events simultaneously assuming a 
physically accurate model.  This model includes dynamical interactions and an accurate 
description of the photometric noise.

In detail, the light curve and radial velocity of Kepler-56 were modeled using a 
dynamical simulation to determine the motions of the planets and star and a transit 
light curve model to predict the light curve at the observed times.  In addition to 
this deterministic model, an extended noise model was fitted to account for the 
significant time-correlated stellar granulation signal superposed with Poisson photon 
noise.  The posterior distribution of the model parameters was sampled using a Markov 
chain Monte Carlo algorithm.  The details of this model, its application, and the 
derived results are described in this section.

\subsection{Preparation of the Light Curve Data}

We isolated the observations near the planetary transit events in the
full {\em Kepler} light curve (specifically ``SAP\_FLUX'') for
Kepler-56.  We retain 96 continuous segments of 256 cadences (roughly
5.2 days) centered on single transit events or, when transits of both
planets occur within 5.2 days of one another, centered halfway between
two transits. We choose 256 cadences --- a power of 2 --- to
facilitate the rapid computation of the wavelet transform when
computing the likelihood (see \S\ref{sec:noise}).  Figures
\ref{fig:datab}, \ref{fig:datac} show a portion of the data utilized
in our analysis, within $\sim1$ day of a transit event.

A quadratic trend in time was fitted to each continuous segment and divided through 
the data.  The parameters of this quadratic trend were found iteratively, re-estimated 
after fitting the data with the photometric-dynamical model using a nonlinear fitter 
(Levenberg-Marquardt).  At each iteration step, the best-fitting light curve model and 
correlated noise model were removed from the data and the quadratic trend was refit to 
the residuals; the revised trend was divided through the data and the process was 
repeated until the parameters of the trend converged to sufficient tolerance.

\begin{figure}
\includegraphics[width=\hsize]{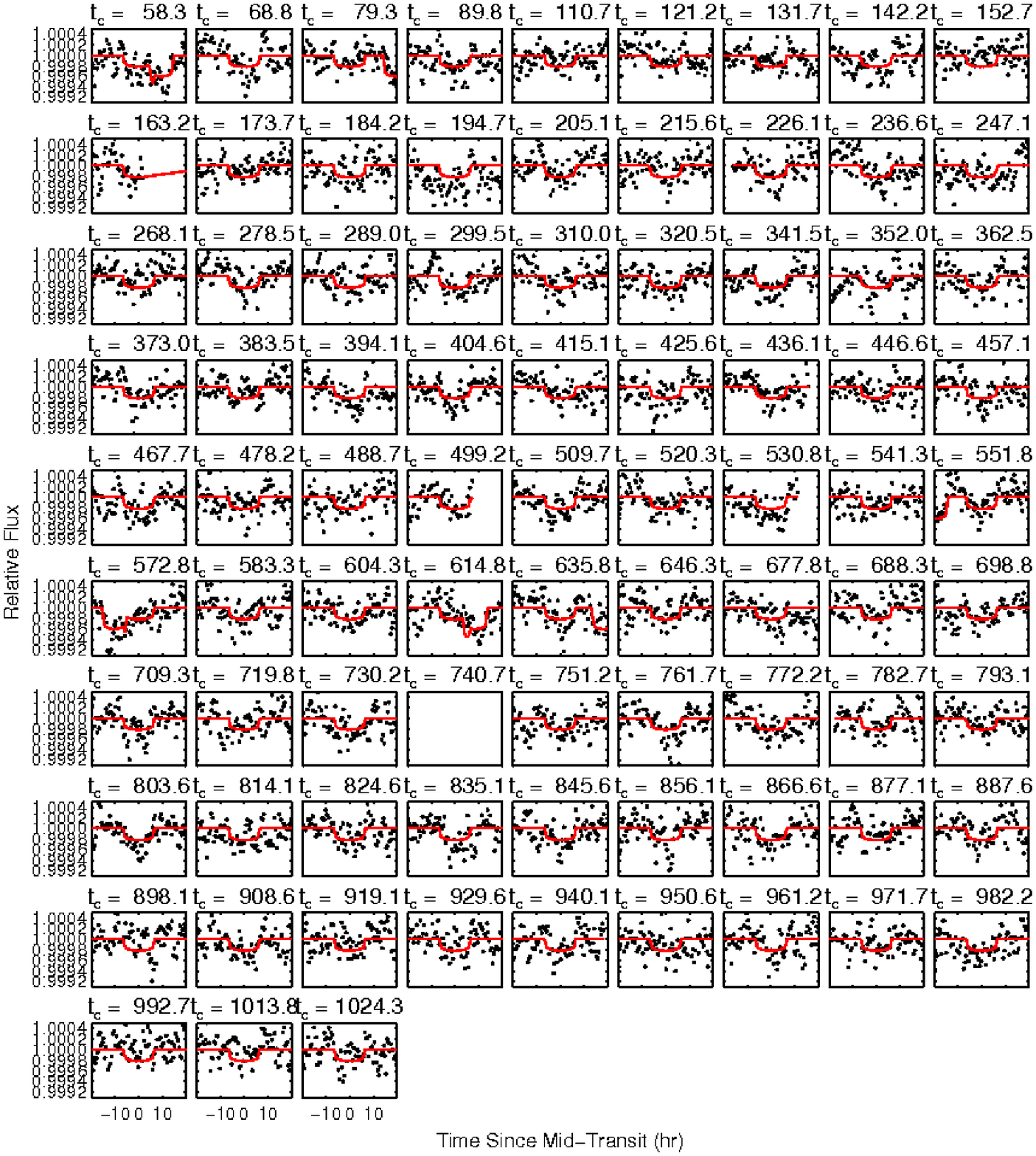}
\caption{ Light curve data and best-fitting transit model (in red) near transits of 
planet b. Times of transit less $2,454,900$ BJD are indicated above each plot cell. 
Note that empty panels are due to data gaps in the Kepler time series. 
\label{fig:datab}}
\end{figure}

\begin{figure}
\includegraphics[width=\hsize]{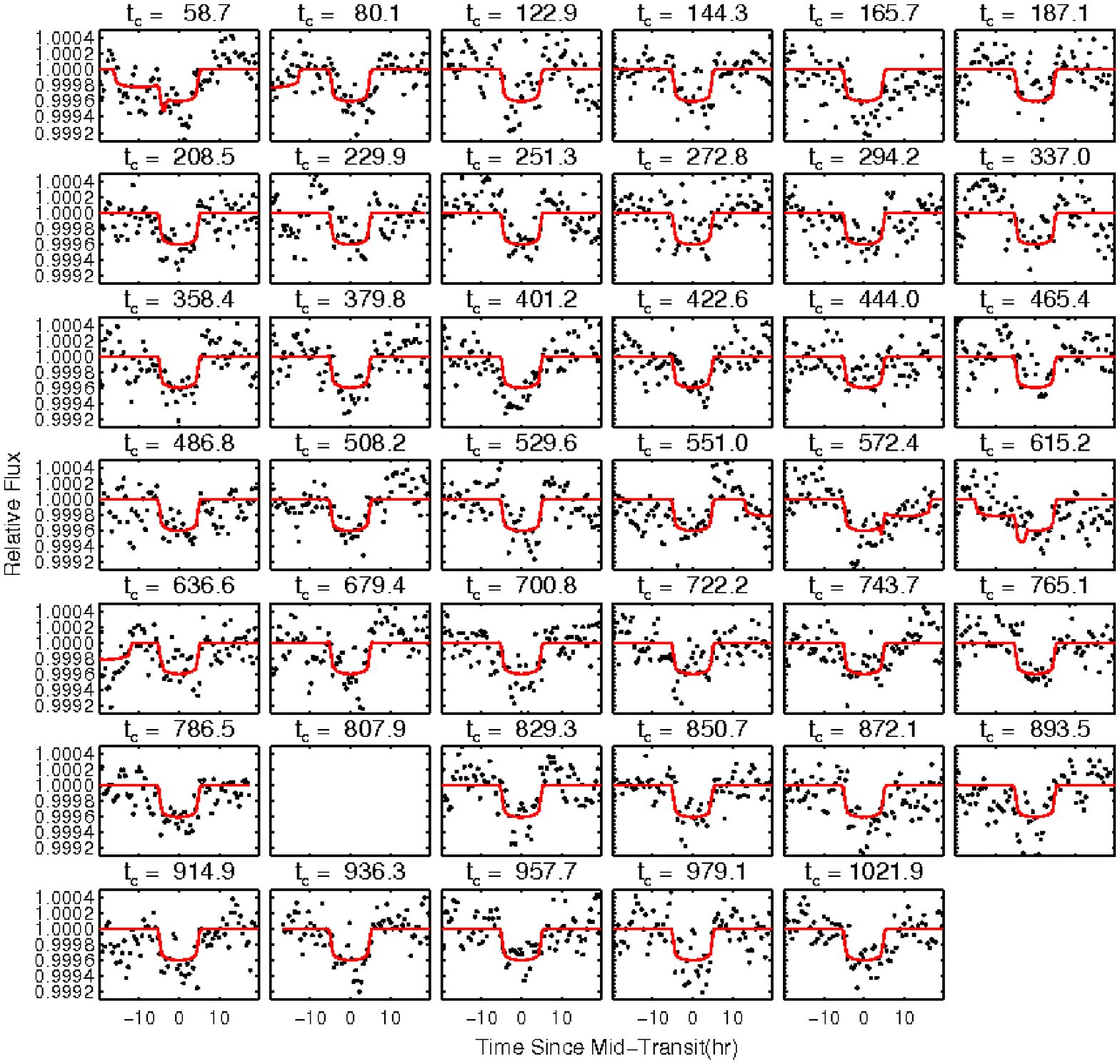}
\caption{ Light curve data and best-fitting transit model (in red) near transits of 
planet c. Times of transit less $2,454,900$ BJD are indicated above each plot cell. 
Note that empty panels are due to data gaps in the Kepler time series.  
\label{fig:datac}}
\end{figure}

\subsection{Dynamical Simulation}

We perform a dynamical integration to determine the positions and velocities of all 
three bodies in Kepler-56 at any time.  This integration utilized a Jacobian coordinate 
system \cite{murray99}.  In this system, ${\bf r_b}$ is the position of planet b 
relative to the star, and ${ \bf r_c}$ is the position of planet c relative to the 
center of mass of planet b and the star.  These coordinates and masses are specified 
(via the parameterization described in \S\ref{sec:params}) at some fiducial time to 
uniquely specify the evolutionary history over our observations. 

The computations are performed in a Cartesian system, although it is
convenient to express ${\bf r_b}$ and ${\bf r_c}$ and their time
derivatives in terms of osculating Keplerian orbital elements:
instantaneous period, eccentricity, argument of pericenter,
inclination, longitude of the ascending node, and time of transit.  We
denote these quantities as $P_{b,c}$, $e_{b,c}$,
$\omega_{b,c}$, $i_{b,c}$, $\Omega_{b,c}$, and $T_{b,c}$, respectively.  We note
that these parameters do not necessarily reflect observables in the
light curve; the unique three-body effects make these parameters
functions of time. The ``time of transit,'' in particular, refers to
the modeled time of transit at the reference epoch; it cannot be used
in conjunction with the modeled orbital period to compute a simple
ephemeris for the system, due to transit timing variations.

The accelerations of the three bodies are determined from Newton's equations of motion, 
which depend on ${\bf r_b}$, ${\bf r_c}$ and the masses \cite{murray99,mardling02}.  
For the purpose of reporting the masses and radii in Solar units, we assumed 
$G \msun = 2.959122 \times 10^{-4}$ AU$^{3}$~day$^{-2}$ and 
$\rsun = 0.00465116$~AU.  We used a Bulirsch-Stoer algorithm \cite{press02} to 
integrate the coupled first-order differential equations for $\dot{\bf r}_{b,c}$ and 
${\bf r}_{b,c}$. We set a positional accuracy of $10^{-16}$~AU.  
The positions and velocities determined from the dynamical simulation were then used as 
input to a light curve model.

\subsection{Light Curve Model}

We did not determine the spatial coordinates of all three bodies at each observed 
time in the {\em Kepler} light curve.  Instead, to speed computation,
we recorded for each epoch only 
the sky-plane projected separation between star and planet, and the sky-plane projected 
speed of planet relative to star at the calculated time of transit.  
The times of transit were determined numerically by minimizing the projected 
separation between the star and planet.  The result of these calculations 
was a collection of transit times $t^{k}_{i_k}$, impact parameters $b^{k}_{i_k}$ and 
speeds $v^{k}_{i_k}$ for each planet $k \in \{b,c\}$ and for epochs $i_k \in N_k$ 
where $N_k$ is the set of observed epoch numbers for planet $k$.  The motion of the 
planet relative to the star is approximately linear in the sky-plane such that the 
projected separation as a function of time is, to good approximation,
\begin{eqnarray}
	Z^k_{i_k}(t) = \sqrt{\left[v^k_{i_k}(t-t^k_{i_k})\right]^2+\left(b^k_{i_k}\right)^2}
\end{eqnarray}
for times near (a few transit durations) of the calculated mid-transit time.

The approximate photometric model for the relative stellar flux, $f(t)$, is then 
defined as
\begin{eqnarray}
	f(t) = 1-\sum_k \sum_{i_k \in N_k} \left\{
		\begin{array}{ll}
			\lambda\left(Z^k_{i_k}(t),R^k,u\right) &  -0.5 \leq t-t^k_{i_k} \leq 0.5 \\
			0 & {\rm otherwise}
		\end{array}
		\right.
\end{eqnarray}
where $\lambda(z,r,u)$ is the overlap integral between a limb darkened star of 
radius $R_\star$ (such that the radial brightness profile is 
$I(\rho/R_\star)/I(0) = 1-u [1-\sqrt{1-(\rho/R_\star)^2}]$ with linear limb-darkening 
parameter $u$) whose center is separated by a distance $z$ from a dark, opaque sphere 
of radius $r$.   $\lambda(z,r,u)$ may be computed semi-analytically with available 
codes \cite{mandel02}.  This photometric model, assuming constant transit 
velocity, is faster to compute than calculating the positions at each photometric 
cadence and results in a negligible change in the quality of the model fit to the 
data compared to exact integration.  This model does not include the ``anomalous'' 
brightening events that occur when the planet c occults planet b during a transit 
\cite{ragozzine10}.  No such events are observed, nor are they
predicted to occur, within the current dataset.

The continuous model $f(t)$ is integrated over a 29.4~min interval
centered on each long cadence sample using a Gaussian-quadrature
integration with 10 samples per cadence.

\subsection{Photometric Noise Model \label{sec:noise}}

Investigation of the {\em Kepler} light curve shows a significant correlated stochastic 
signal in addition to the coherent oscillations (utilized in the asteroseismic analysis) 
and the transit events (see Figure~\ref{fig:correxamp}).  This signal is attributed to 
stellar granulation, is approximately stationary (i.e., temporal correlation depends 
only on relative separations in time), and has a power spectral density that scales 
inversely with frequency. This nearly $1/f$ (``pink'') power spectral
density is typical 
of granulation noise \cite{press78,kjeldsen11,mathur11b}.  This correlated noise can significantly 
bias parameters related to the transit events (e.g., mid-transit times, depths) if not 
properly accounted for \cite{carter09}.   

In response, we model the photometric noise, $\eta$, as $\eta(t) = \epsilon+\gamma(t)$ 
where $\epsilon$ and $\gamma$ are both normally distributed 
[$\epsilon \sim {\cal N}(0; \Sigma_w)$, $\gamma \sim {\cal N}(0; \Sigma_p)$], $\epsilon$ 
is uncorrelated ``white'' noise, and $\gamma$ is correlated ``pink'' noise. To facilitate 
the rapid computation of this model and its associated likelihood, we use the wavelet-based 
formalism described in Carter \& Winn (2009) \cite{carter09}.  Here, the data 
$\eta(t)$ (or the data residuals after removing the transit model, $\eta(t) = F(t)-f(t)$) 
are projected into components $\hat{\eta}_{m,n}$ of a wavelet basis (indexed by scale $m$ 
and position $n$).  In this basis, the covariance of the components is approximately 
diagonal
\begin{eqnarray}
\langle \hat{\eta}_{m,n} \hat{\eta}_{m',n'} \rangle &\approx & (\sigma_r^2 2^{-m}+\sigma_w^2) \delta_{m,m'} \delta_{n,n'}
\end{eqnarray}
where we have parameterized the noise model by two parameters, $\sigma_w^2$ and $\sigma_r^2$.  
The first parameter is the variance of $\epsilon$ (associated with photon-noise only) while 
the second parameter is related to the scale of the correlated noise component. 

\begin{figure}
\includegraphics[width=\hsize]{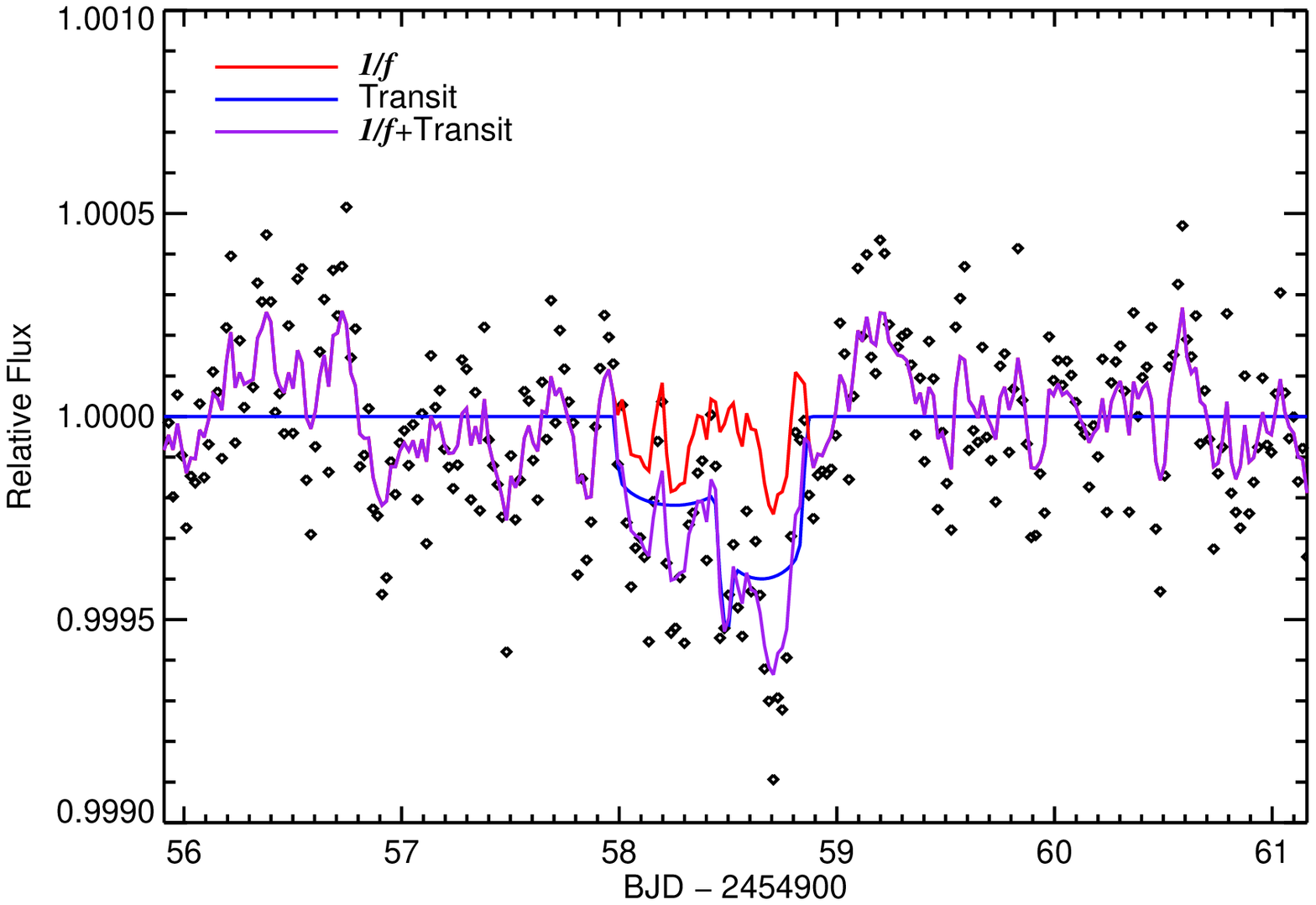}
\caption{Example decomposition of single data segment into its constitute components: 
transit model (blue; itself the superposition of a nearly simultaneous transits of 
Planets b and c) and a stochastic correlated noise component (red).  
\label{fig:correxamp}}
\end{figure}

\subsection{Specification of Parameters \label{sec:params}}

The reference epoch of the initial conditions was chosen to be $t_0 = 2,454,950$ (BJD).
The photo-dynamical model has 23 adjustable parameters. Two parameters are related to stellar 
constraints from asteroseismology: the stellar density times the gravitational 
constant, $G \rho_\star$, and the stellar radius, $R_\star$. Two parameters are the 
mass ratios $q_+ \equiv (M_b+M_c)/M_\star$ and $q_p \equiv M_b/M_c$. Four parameters 
are combinations of the eccentricities $e_{b,c}$ and arguments of pericenter $\omega_{b,c}$ 
in a nonlinear way, chosen to give nearly linear correlations between their
uncertainties
(and thereby avoid the computational cost often associated with nonlinear
correlations):
\begin{eqnarray}
	h_- &\equiv & (P_b/P_c)^{2/3}  e_b\cos \omega_b-e_c \cos \omega_c \\
	h_+ & \equiv & (P_b/P_c)^{2/3}  e_b\cos \omega_b+e_c \cos \omega_c\\
	k_- & \equiv &(P_b/P_c)^{2/3}  e_b\sin \omega _b- e_c \sin \omega_c \\
	k_+ & \equiv & (P_b/P_c)^{2/3}  e_b \sin \omega_b + e_c \sin \omega_c 
\end{eqnarray}

The remaining osculating parameters, 7 in total, are the periods $P_b$, $P_c$, the 
orbital inclinations $i_b$, $i_c$, the times of transit $T_b$, $T_c$ and the difference 
between the nodal longitudes $\Delta\Omega \equiv \Omega_c-\Omega_b$. 

Two more parameters are the relative radii of the planets: $r_b \equiv R_b/R_\star$ 
and $r_c \equiv R_c/R_\star$.  One parameter, $u$, parameterizes the linear limb 
darkening law for the star.  

Two parameters, $\sigma_w^2$ and $\sigma_r^2$, characterize the extended noise model 
(see \S~\ref{sec:noise}).

The remaining parameters parameterize the radial velocity model: 3 describe the 
quadratic trend and one parameter gives the additional stellar jitter 
($\sigma_{\rm jitter}$), added in quadrature to the formal velocity errors. 

\subsection{Model Likelihood and Priors}

We adopted uniform priors in the parameters described in the previous section 
excluding $h_{+,-}$ and $k_{+,-}$.  For these latter four parameters, we enforced 
uniform priors in eccentricities and arguments of pericenter.  
For these priors, the probability density obeys
\begin{eqnarray}
	p( h_{+,-}, k_{+,-}) d h_{+,-}  k_{+,-} &\propto& p(e_{b,c},\omega_{b,c}) \times  \frac{1}{e_b e_c} d_{e_b,e_c} d_{\omega_b, \omega_c} \propto \frac{1}{e_b e_c} d_{e_b,e_c} d_{\omega_b, \omega_c}
\end{eqnarray}

The likelihood ${\cal L}$ of a given set of parameters was taken to be the product of 
likelihoods based on the photometric data (each 256 cadence segment projected into a 
discrete fourth-order Daubechies wavelet basis), the assumed-Gaussian asteroseismology 
priors and the radial velocity data: 
\begin{eqnarray}
	{\cal L} &\propto& \prod^{{\rm segments}}\left[\prod_{n} \prod_{m} (\sigma_r^2 2^{-m}+\sigma_w^2)^{-\frac{1}{2}} \exp\left(-\frac{1}{2}\frac{\hat{\eta}^s_{m,n}}{\sigma_r^2 2^{-m}+\sigma_w^2}\right)\right] \\ \nonumber
		&& \times \prod_i (\sigma_i^2+\sigma_{\rm jitter}^2)^{-\frac{1}{2}} \exp\left(-\frac{1}{2}\frac{\Delta RV_i^2}{\sigma_i^2+\sigma_{\rm jitter}^2}\right) \times \\ \nonumber
				&& \times \exp \left[-\frac{1}{2}\left(\frac{\Delta G\rho_\star}{\sigma_{G\rho_\star}}\right)^2\right] \times \exp \left[-\frac{1}{2}\left(\frac{\Delta R_\star}{\sigma_{R_\star}}\right)^2\right] \nonumber 
\end{eqnarray}
where $\hat{\eta}^s_{m,n}$ are the wavelet components of the $s$th segment photometric 
residuals after removing the transit model $f(t)$ (see Carter \& Winn 2009 for 
additional details), $\Delta RV_i$ is the residual of the $i$th radial velocity 
measurement with formal error $\sigma_i$, and $\Delta G\rho_\star/\sigma_{G\rho_\star}$ 
and $\Delta R_\star/\sigma_{R_\star}$ are the deviates between the asteroseismic 
constraints in density and radius.

\subsection{Parameter Estimation}

We explored the parameter space and estimated the posterior parameter distribution with 
a Differential Evolution Markov Chain Monte Carlo (DE-MCMC) algorithm \cite{braak06}. 
We generated a population of 60 chains and evolved through approximately 500,000 
generations.  The initial parameter states of the 60 chains were randomly selected 
from an over-dispersed region in parameter space bounding the final posterior 
distribution.  The first 10\% of the links in each individual Markov chain were clipped, 
and the resulting chains were concatenated to form a single Markov chain, after having 
confirmed that each chain had converged according to the standard criteria including 
the Gelman-Rubin convergence statistics and the observation of a long effective chain 
length in each parameter (as determined from the chain autocorrelation).

\subsection{Photodynamical Modeling Results}

Initially, we included only the photometric data subject to the asteroseismic 
constraints on stellar density and radius in our analysis, excluding the radial 
velocity data.   Examining the MCMC results, we found that at low mutual orbital 
inclination $I$, defined such that
\begin{eqnarray}
	\cos I & = & \sin i_b \sin i_c \cos \Delta \Omega + \cos i_c \cos i_b,
\end{eqnarray}
the planetary orbits were nearly circular and the planetary masses were moderately 
constrained (to within $\approx$10\%).

Arbitrarily high mutual inclinations are marginally consistent (Figure~\ref{fig:ll}) 
with the photometric data (in the tail of the posterior distribution) so long as the 
planetary masses, orbital eccentricities and arguments of periapse are relatively 
fine-tuned.  These dependences are shown in Figure~\ref{fig:del}.  The source of these 
curious correlations is the changing character of the transit timing anomalies at mutual 
inclinations exceeding roughly 20 degrees.  

\begin{figure}
\begin{center}
\includegraphics[width=6.in]{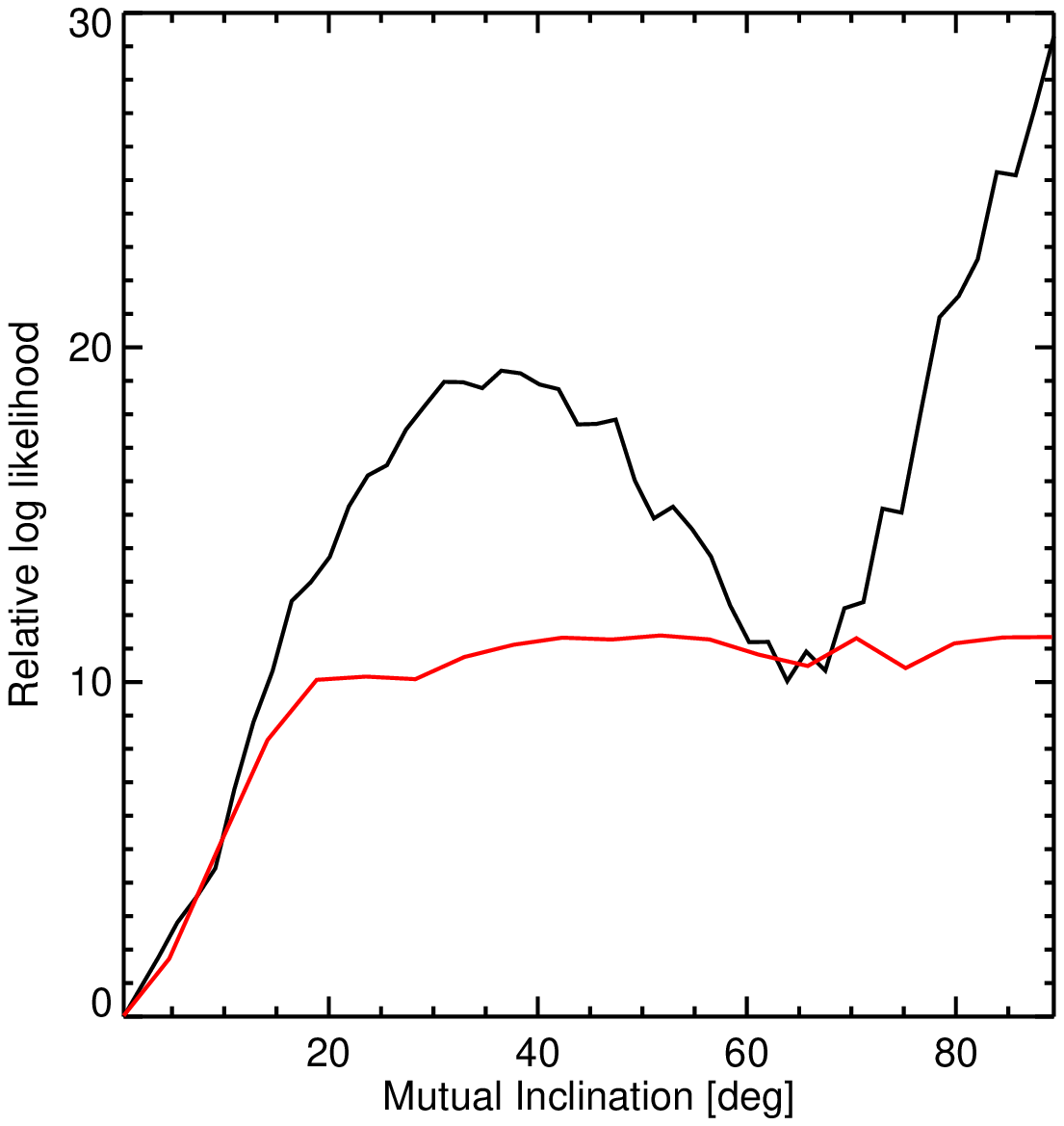}
\caption{ Relative log-likelihood as a function of mutual inclination estimated from 
the photo-dynamical model when RV data 
is excluded (red) or included (black).   Smaller values correspond to higher likelihood.     
\label{fig:ll}}
\end{center}
\end{figure}

\begin{figure}
\begin{center}
\includegraphics[width=5.in]{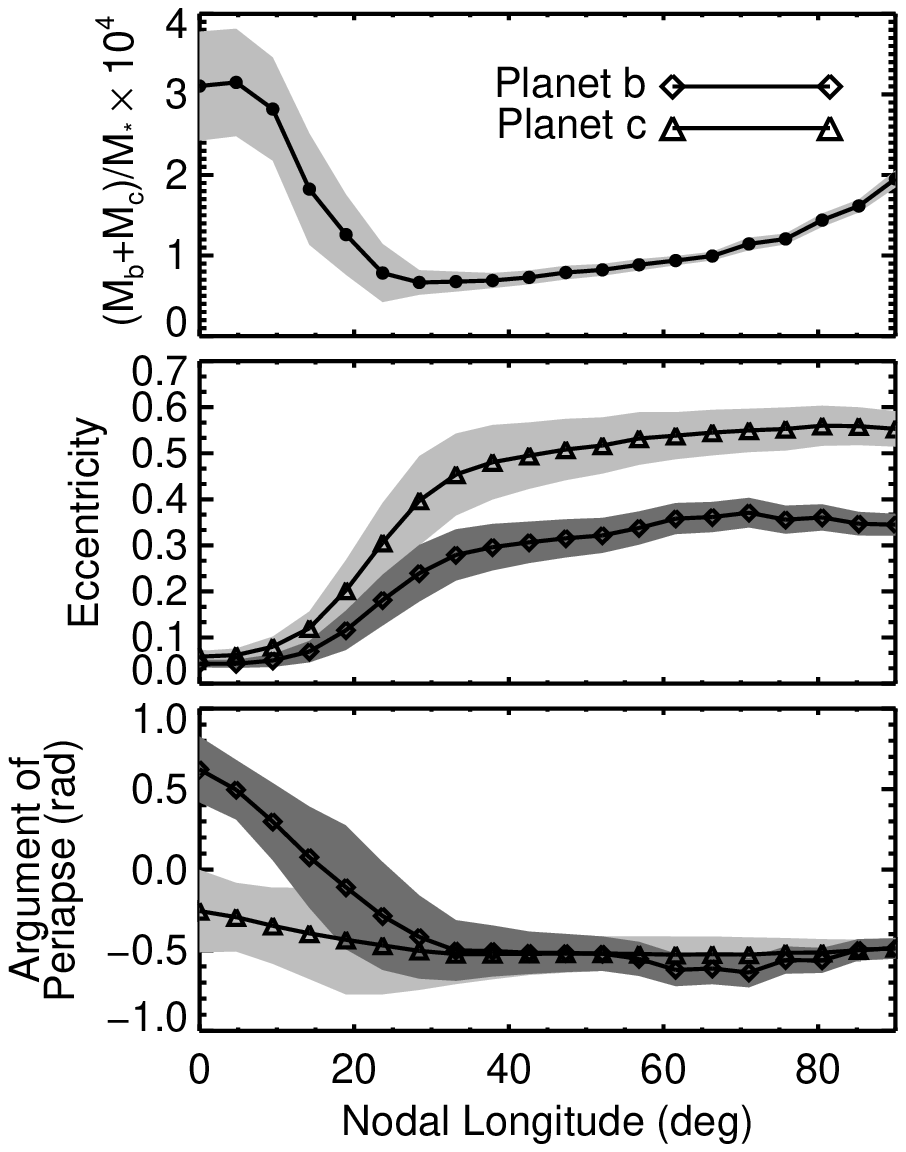}
\caption{  Best-fitting parameters as a function of the relative nodal longitude between 
the orbits of Planet b and c, fitting only the photometric data (subject to the constraints 
from asteroseismology).   The shaded gray regions indicate the 1$\sigma$ intervals of 
uncertainty. The best-fitting architecture undergoes a finely-tuned ``phase change'' 
for mutual inclinations (approximately equal to the relative nodal longitude) greater 
than 20$^\circ$.   \label{fig:del}}
\end{center}
\end{figure}

At low mutual inclinations, the periodicity of the timing anomaly is determined by the 
period of the longitude of conjunction, defined as the mutual anomaly 
at planetary conjunction \cite{agol05,lithwick12}: $P_{\rm LOC} = |2/P_c - 1/P_b|^{-1}$.
At high inclinations, additional minima in 
separation appear near conjunctions at either node (orbital plane crossings) and occur 
twice each $P_{\rm LOC}$.  As a result, a frequency doubled component arises in the TTV 
at high inclinations (see Figure~\ref{fig:freqdouble}).  The onset of this frequency 
doubling occurs at moderate mutual inclinations.  

\begin{figure}
\begin{center}
\includegraphics[width=5in]{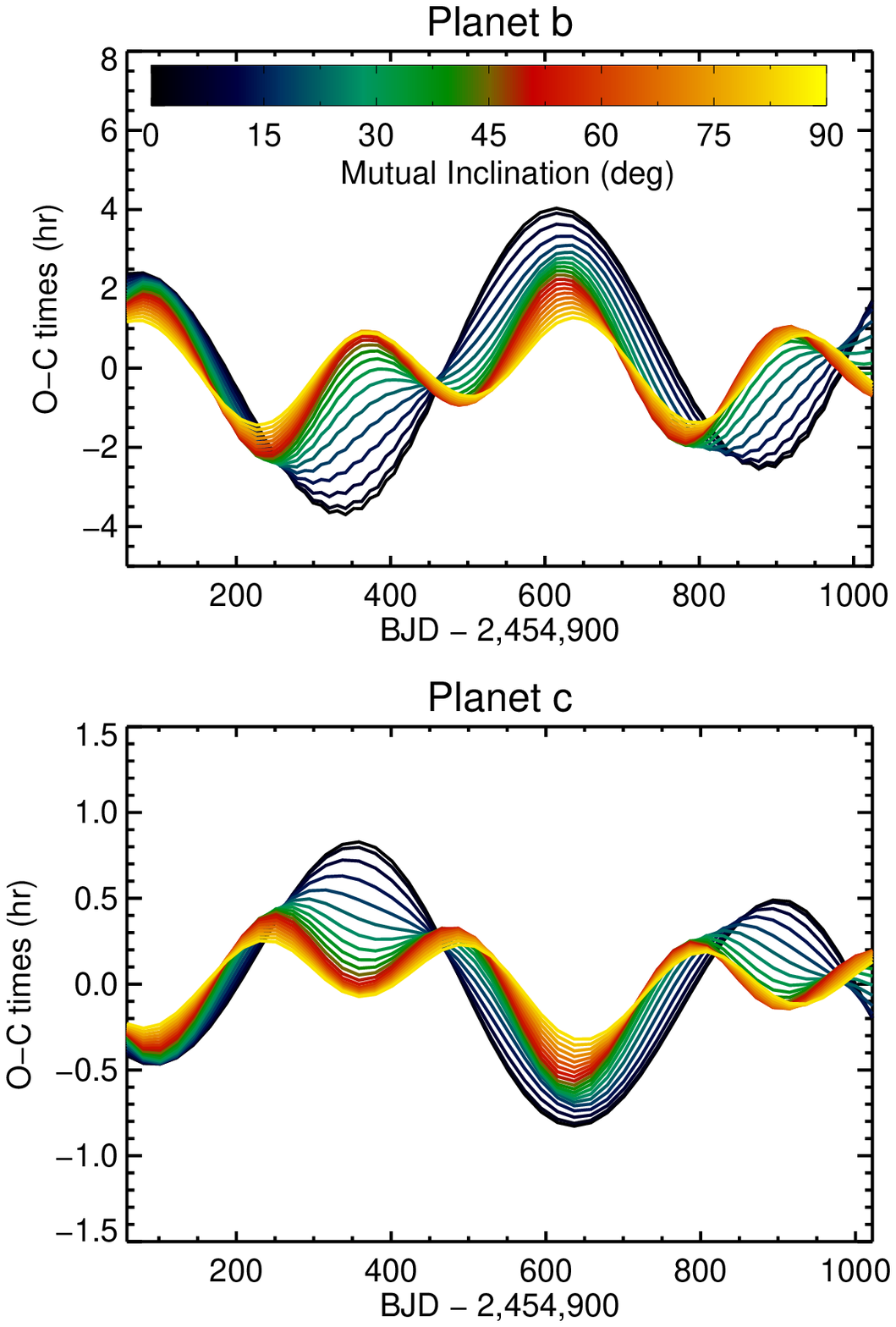}
\caption{ Transit timing variation for Kepler-56 b and c as a function of mutual inclination 
keeping all other parameters fixed. \label{fig:freqdouble}}
\end{center}
\end{figure}

The data favor a single periodicity (see Figure 2) at the expected
period $P_{\rm LOC} \approx 590$ days \cite{lithwick12}. However, by
carefully orienting the orbits and increasing eccentricities (and
finely tuning their masses to keep the TTV amplitude constant), the
frequency doubled component can be suppressed.  In particular, when
pericenters are aligned and the eccentricity is sufficiently large the
longitude of conjunction sweeps quickly through the line of nodes; in
this case, the behavior of the TTV is approximately described with a
single component.  This is demonstrated in the correlations shown in
Figure~\ref{fig:del} for mutual inclinations $I > 20$ degrees; these
high eccentricity solutions have approximately equal likelihoods,
slightly lower than low inclination solutions (see
Figure~\ref{fig:ll}).


The inclusion of the radial velocity data in our analysis resolved
this degeneracy between mutual inclination and planet mass sum or
orbital eccentricity.  In detail, the radial velocity data constrained
both the planetary masses and the eccentricity of their orbits to a
range consistent with low mutual inclination.  Moderate mutual
inclinations ($15^\circ < I < 50^\circ$) and high mutual inclinations
($I > 75^\circ$) were excluded by the data. A lower likelihood
connected region near $I \approx 60^\circ$ was statistically plausible
(Figure~\ref{fig:ll}), but was ultimately excluded based on
considerations of long-term dynamical stability (see \S\ref{sec:dyn}).
We conclude that the orbits are coplanar to within $I < 10^\circ$ at
95\% confidence (Figure~\ref{fig:mut}).

\begin{figure}
\begin{center}
\includegraphics[width=\hsize]{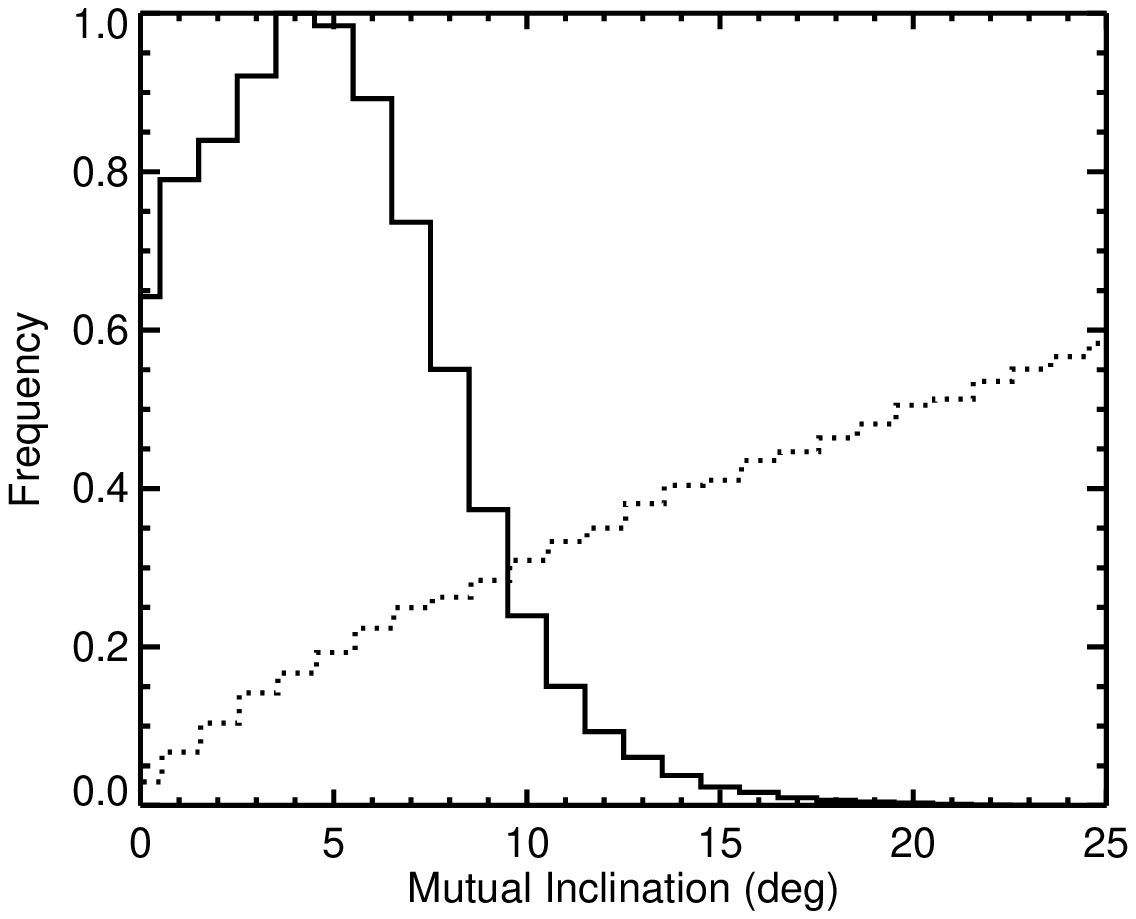}
\caption{Distribution of mutual inclination, $I$, determined from the photometric-dynamical 
analysis (solid line) and the prior for $I$ (dotted line). Both distributions have 
been normalized so that the maximum equals unity.
\label{fig:mut}}
\end{center}
\end{figure}

Table \ref{tab:parameters} provides the best-fitting photometric-dynamical model 
parameters, their medians and uncertainties (provided as 68\% confidence intervals 
of the marginalized parameter posterior drawn by the MCMC algorithm).  
Table~\ref{tab:derived} provides a number of derived parameters.  Figure~\ref{fig:corr} 
shows 2D joint probability distribution plots between the fitted parameters (and 
references the parameter indices listed in Table~\ref{tab:parameters}).  These plots are 
meant to qualitatively display the correlations amongst the parameters.

 \begin{table}
 \begin{small}
\centering
\begin{tabular}{|ll|l|lll|}
\hline
Index & Parameter Name & Best-fit & 50\% & 15.8\% & 84.2\% \\ \hline
&~{\it Mass parameters} & & & & \\
0&~~Mean Density, $\rho_\star$ (g cm$^{-3}$) & $                                 0.02458$ & $                                 0.02461 $ & $-                                 0.00060$ & $+                                 0.00059$\\
1&~~Mass sum ratio, $(M_b+M_c)/M_A (\times 10^5)$ & $                                    47.7$ & $                                    46.8 $ & $-                                     3.9$ & $+                                     3.9$\\
2&~~Planetary mass ratio, $M_b/M_c$ & $                                   0.129$ & $                                   0.122 $ & $-                                   0.015$ & $+                                   0.015$\\
&~{\it Inner Binary Orbit (Planet b)} & & & &\\
3&~~Orbital Period, $P_b$ (day) & $  10.51046$ & $  10.51057 $ & $-                                  0.0010$ & $+                                  0.0011$\\
4&~~Time of Transit, $t_b$ (days since $t_0$) & $                                  8.2581$ & $                                  8.2556 $ & $-                                  0.0057$ & $+                                  0.0056$\\
5&~~Orbital Inclination, $i_b$ (deg) & $  83.84$ & $  83.92 $ & $-                                    0.25$ & $+                                    0.26$\\
&~{\it Outer Binary Orbit (Planet c)} & & & &\\
6&~~Orbital Period, $P_c$ (day) & $  21.40221$ & $  21.40239 $ & $-                                 0.00062$ & $+                                 0.00059$\\
7&~~Time of Transit, $t_c$ (days since $t_0$) & $                                  8.6531$ & $                                  8.6560 $ & $-                                  0.0055$ & $+                                  0.0057$\\
8&~~Orbital Inclination, $i_c$ (deg) & $ 84.02$ & $ 84.08 $ & $-                                   0.087$ & $+                                   0.091$\\
9&~~Relative Nodal Longitude, $\Delta \Omega$ (deg) & $ -4.91$ & $ -4.95 $ & $-                                     3.5$ & $+                                     3.8$\\
&~{\it Eccentricity parameters} & & & & \\
10&~~$e_b \cos \omega_b-(a_c/a_b) e_c \cos \omega_c$ & $                                   0.033$ & $                                   0.032 $ & $-                                   0.022$ & $+                                   0.018$\\
11&~~$e_b \cos \omega_b+(a_c/a_b) e_c \cos \omega_c$ & $                                   0.033$ & $                                   0.034 $ & $-                                   0.023$ & $+                                   0.032$\\
12&~~$e_b \sin \omega_b-(a_c/a_b) e_c \sin \omega_c$ & $                                  -0.010$ & $                                  -0.004 $ & $-                                   0.013$ & $+                                   0.019$\\
13&~~$e_b \sin \omega_b+(a_c/a_b) e_c \sin \omega_c$ & $                                  -0.010$ & $                                  -0.020 $ & $-                                   0.031$ & $+                                   0.018$\\
&~{\it Radius Parameters} & & & &\\
14&~~Linear Limb Darkening Parameter, $u$ & $                                   0.464$ & $                                   0.530 $ & $-                                   0.100$ & $+                                   0.091$\\
15&~~Stellar Radius, $R_\star$ ($R_\odot$) & $                                    4.19$ & $                                    4.22 $ & $-                                    0.15$ & $+                                    0.15$\\
16&~~b Radius Ratio, $R_b/R_\star$ & $                                 0.01419$ & $                                 0.01414 $ & $-                                 0.00038$ & $+                                 0.00037$\\
17&~~c Radius Ratio, $R_c/R_\star$ & $                                 0.02109$ & $                                 0.02130 $ & $-                                 0.00065$ & $+                                 0.00064$\\
&~{\it Photometric Noise Parameters} & & & & \\
18&~~White Noise paramter, $\sigma_{w}$ ($\times 10^5$) & $                                   12.78$ & $                                   12.80 $ & $-                                    0.20$ & $+                                    0.19$\\
19&~~Pink Noise parameter, $\sigma_{r}$ ($\times 10^5$) & $                                   152.3$ & $                                   152.5 $ & $-                                     2.3$ & $+                                     2.3$\\
&~{\it RV Parameters} & & & & \\
20&~~RV Offset (m/s) & $                                    13.1$ & $                                    13.0 $ & $-                                     2.6$ & $+                                     2.7$\\
21&~~Linear Trend (m/s/day) & $                                    0.79$ & $                                    0.86 $ & $-                                    0.12$ & $+                                    0.12$\\
22&~~Quadratic Trend (m/s/day$^2$) & $                                  0.0017$ & $                                  0.0028 $ & $-                                  0.0025$ & $+                                  0.0024$\\
23&~~RV Jitter (m/s) & $                                     5.2$ & $                                     5.9 $ & $-                                     1.7$ & $+                                     2.8$\\
 \hline
\end{tabular}
\caption{ Photometric-dynamical model parameters. The reference epoch is 
$t_0 = $2,454,950 (BJD). \label{tab:parameters}}
 \end{small}
\end{table}

 \begin{table}
  \begin{small}
\centering
\begin{tabular}{|ll|l|ll|}
\hline
~~Parameter & Best-fit & 50\% & 15.8\% & 84.2\% \\ \hline
~{\it Planetary Bulk Properties} & & & & \\
~~Mass of Planet b, $M_b$ ($M_\oplus$) & $                                    23.1$ & $                                    22.1 $ & $-                                     3.6$ & $+                                     3.9$\\
~~Mass of Planet c, $M_c$ ($M_\oplus$) & $                                    180.$ & $                                    181. $ & $-                                     19.$ & $+                                     21.$\\
~~Radius of Planet b, $R_b$ ($R_\oplus$) & $                                    6.47$ & $                                    6.51 $ & $-                                    0.28$ & $+                                    0.29$\\
~~Radius of Planet c, $R_c$ ($R_\oplus$) & $                                    9.63$ & $                                    9.80 $ & $-                                    0.45$ & $+                                    0.46$\\
~~Density of Planet b, $\rho_b$ (g cm$^{-3}$) & $                                   0.468$ & $                                   0.442 $ & $-                                   0.072$ & $+                                   0.080$\\
~~Density of Planet c, $\rho_c$ (g cm$^{-3}$) & $                                    1.11$ & $                                    1.06 $ & $-                                    0.13$ & $+                                    0.14$\\
~~Planetary Density Ratio, $\rho_b/\rho_c$ & $                                   0.423$ & $                                   0.417 $ & $-                                   0.066$ & $+                                   0.075$\\
~~Planet b to Star Density Ratio, $\rho_b/\rho_\star$ & $                                    19.1$ & $                                    18.0 $ & $-                                     2.9$ & $+                                     3.2$\\
~~Planet c to Star Density Ratio, $\rho_c/\rho_\star$ & $                                    45.0$ & $                                    43.1 $ & $-                                     5.1$ & $+                                     5.7$\\
~~Surface Gravity of Planet b, $g_b$ (m s$^{-2}$) & $                                    5.40$ & $                                    5.13 $ & $-                                    0.78$ & $+                                    0.84$\\
~~Surface Gravity of Planet c, $g_c$ (m s$^{-2}$) & $                                    19.0$ & $                                    18.5 $ & $-                                     1.8$ & $+                                     1.9$\\
~~Escape Velocity of Planet b, $v_{{\rm esc},b}$ (km s$^{-1}$) & $                                    21.1$ & $                                    20.6 $ & $-                                     1.6$ & $+                                     1.6$\\
~~Escape Velocity of Planet c, $v_{{\rm esc},c}$ (km s$^{-1}$) & $                                    48.3$ & $                                    48.1 $ & $-                                     2.3$ & $+                                     2.3$\\
~{\it Orbital Properties} & & & & \\
~~Semimajor Axis of Planet b, $a_b$ (AU) & $                                  0.1019$ & $                                  0.1028 $ & $-                                  0.0037$ & $+                                  0.0037$\\
~~Semimajor Axis of Planet c, $a_c$ (AU) & $                                  0.1637$ & $                                  0.1652 $ & $-                                  0.0059$ & $+                                  0.0059$\\
~~Mutual Orbital Inclination, $I$ (deg) & $                                     4.9$ & $                                     5.0 $ & $-                                     3.1$ & $+                                     3.4$\\
~~Orbital Velocity of Planet b, $2\pi a_b/P_b$ (km s$^{-1}$) & $                                   105.5$ & $                                   106.4 $ & $-                                     3.8$ & $+                                     3.8$\\
~~Orbital Velocity of Planet c, $2\pi a_c/P_c$ (km s$^{-1}$) & $                                    83.2$ & $                                    84.0 $ & $-                                     3.0$ & $+                                     3.0$\\
~~Mutual Hill Radius, $R_H$, $\left(\frac{q_+}{24}\right)^{1/3}\left(a_b+a_c\right)$ (AU) & $                                 0.00719$ & $                                 0.00720 $ & $-                                 0.00027$ & $+                                 0.00027$\\
~{\it Transit Parameters} & & & & \\ 
~~Radius Ratio of Planet b, $R_b/R_\star$ & $                                 0.01419$ & $                                 0.01414 $ & $-                                 0.00038$ & $+                                 0.00037$\\
~~Radius Ratio of Planet c, $R_c/R_\star$ & $                                 0.02109$ & $                                 0.02130 $ & $-                                 0.00065$ & $+                                 0.00064$\\
~~Impact Parameter of Planet b, $b_b/R_\star$ & $                                   0.562$ & $                                   0.554 $ & $-                                   0.021$ & $+                                   0.020$\\
~~Impact Parameter of Planet c, $b_c/R_\star$ & $                                  0.8754$ & $                                  0.8673 $ & $-                                  0.0099$ & $+                                  0.0081$\\
~~Transit Velocity of Planet b, $v_b/R_\star$ (day$^{-1}$) & $                                   3.100$ & $                                   3.092 $ & $-                                   0.030$ & $+                                   0.029$\\
~~Transit Velocity of Planet c, $v_c/R_\star$ (day$^{-1}$) & $                                   2.468$ & $                                   2.454 $ & $-                                   0.040$ & $+                                   0.034$\\
~~Transit Duration of Planet b (hr) & $                                   13.08$ & $                                   13.19 $ & $-                                    0.17$ & $+                                    0.17$\\
~~Transit Duration of Planet c (hr) & $                                   10.22$ & $                                   10.53 $ & $-                                    0.33$ & $+                                    0.45$\\
~~Transit Ingress/Egress Duration of Planet b (min) & $                                   15.93$ & $                                   15.82 $ & $-                                    0.58$ & $+                                    0.59$\\
~~Transit Ingress/Egress Duration of Planet c (min) & $                                    51.1$ & $                                    50.3 $ & $-                                     1.9$ & $+                                     1.9$\\
~~Temperature Scaling of Planet b, $\sqrt{R_\star/2a_b}$ & $                                  0.3091$ & $                                  0.3090 $ & $-                                  0.0012$ & $+                                  0.0013$\\
~~Temperature Scaling of Planet c, $\sqrt{R_\star/2a_c}$ & $                                 0.24386$ & $                                 0.24382 $ & $-                                 0.00097$ & $+                                 0.00100$\\
 \hline
\end{tabular}
\caption{Derived parameters. The reference epoch is $t_0 = $2,454,950 (BJD). 
\label{tab:derived}}
 \end{small}
\end{table}

\begin{figure}
\begin{center}
\includegraphics[width=\hsize]{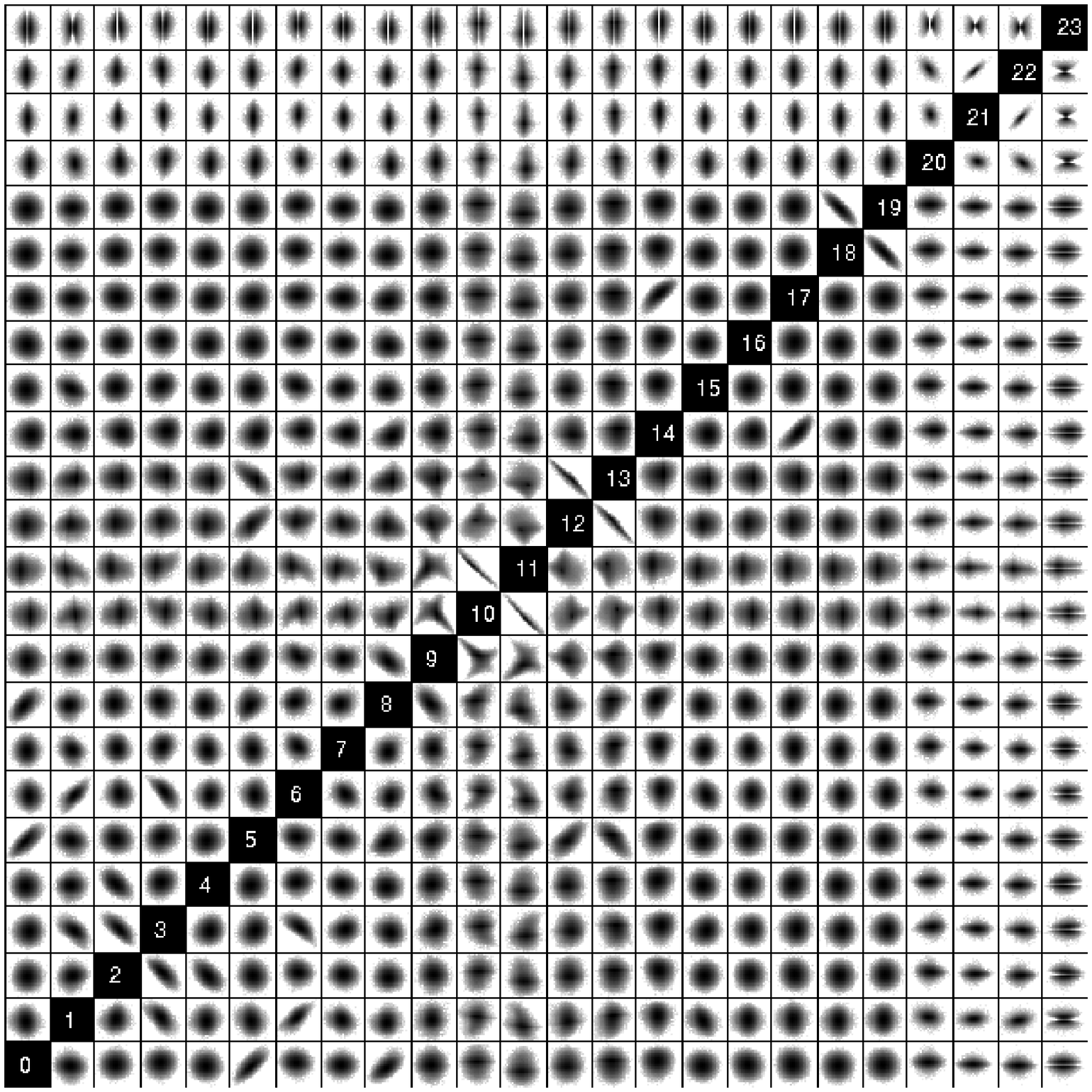}
\caption{ Two-parameter joint posterior distributions of the primary model parameters.  
The densities are plotted logarithmically in order to elucidate the nature of the parameter 
correlations.  The indices listed along the diagonal indicate which parameter is associated 
with the corresponding row and column.  The parameter name corresponding to a given index 
is indicated in Table \ref{tab:parameters} in the ``Index'' column.    \label{fig:corr}}
\end{center}
\end{figure}

\section{Dynamical Stability Analysis of the Inner Planets}
\label{sec:dyn}

The photodynamical modeling of the \id\ data results in a posterior joint 
probability distribution for the model parameters. Marginalizing over all parameters 
except the radius and density of the star (resulting in the mass of the star), the 
two mass ratio parameters and the initial positions and velocities of the bodies at 
a reference epoch results in a posterior distribution for these {\it dynamical} parameters. 
Sets of masses and initial conditions drawn from this distribution are statistically 
consistent with the photometric and radial velocity data. However, there is no guarantee 
that these orbits will be stable on longer timescales. We should reject any initial 
conditions that show instability on timescales much shorter than the age of the system, 
even if they are consistent with the data, unless there is some reason to believe we are 
observing the system at a special time.

\subsection{Orbital Solutions with High Mutual Inclinations}

A set of 6,600 initial conditions, drawn from a Markov chain specially
seeded in the region of parameter space corresponding to highly
inclined solutions near $I \sim 60^{\circ}$, were tested for dynamical stability. All of these
initial conditions failed the Hill stability criterion, implying that
crossing orbits and collisions were possible
\cite{Gladman,Marchal}. Although orbits satisfying the Hill criterion
can never result in particularly strong gravitational interactions (occurring when the 
planets pass within a mutual Hill sphere of each other, for example) or direct collisions
between the bodies, failing the criterion is not sufficient to conclude that
collisions will definitely occur. This is especially true when orbits
are protected by a resonance.  Hence we cannot immediately conclude
that these 6,600 initial conditions are unstable without direct
numerical integration.

The dynamical evolution of these initial conditions was studied by evolving the Newtonian 
equations of motion using a Bulirsch-Stoer integration scheme \cite{BS}. No relativistic 
or dissipative forces were included in these integrations.  Energy was conserved in these 
integrations to within one part in $10^{10}$. 

The integrations revealed that the orbits with high mutual inclinations exhibited large 
amplitude Kozai-like oscillations in the inclinations and eccentricities of the planets.
Though the initial eccentricities of the planets are not particularly large 
($e_b \sim 0.12, e_c \sim 0.21$), they correspond to a minimum in the eccentricity 
oscillations. The maximum eccentricity of the inner planet can approach unity. The 
eccentricity and sky plane inclination evolution is shown in Figure \ref{fig:ecc_ev}. 
Throughout this evolution, the semimajor axes are approximately constant, with only a 
small variation due to coherent oscillations related to the near 2:1 commensurability. 
Because of this, the large growth in eccentricity directly implies that the minimum 
pericenter distance of the inner planet becomes very small.
When the 
eccentricity of the inner planet is larger than 0.8, the corresponding pericenter is 
inside of the host star ($1-e_b< R_\star/a_b = 0.2$, where $a_b$ is roughly constant). 
A typical case is shown in Figure \ref{fig:peri}.

\begin{figure}
\begin{center}
\includegraphics[width=4.5in]{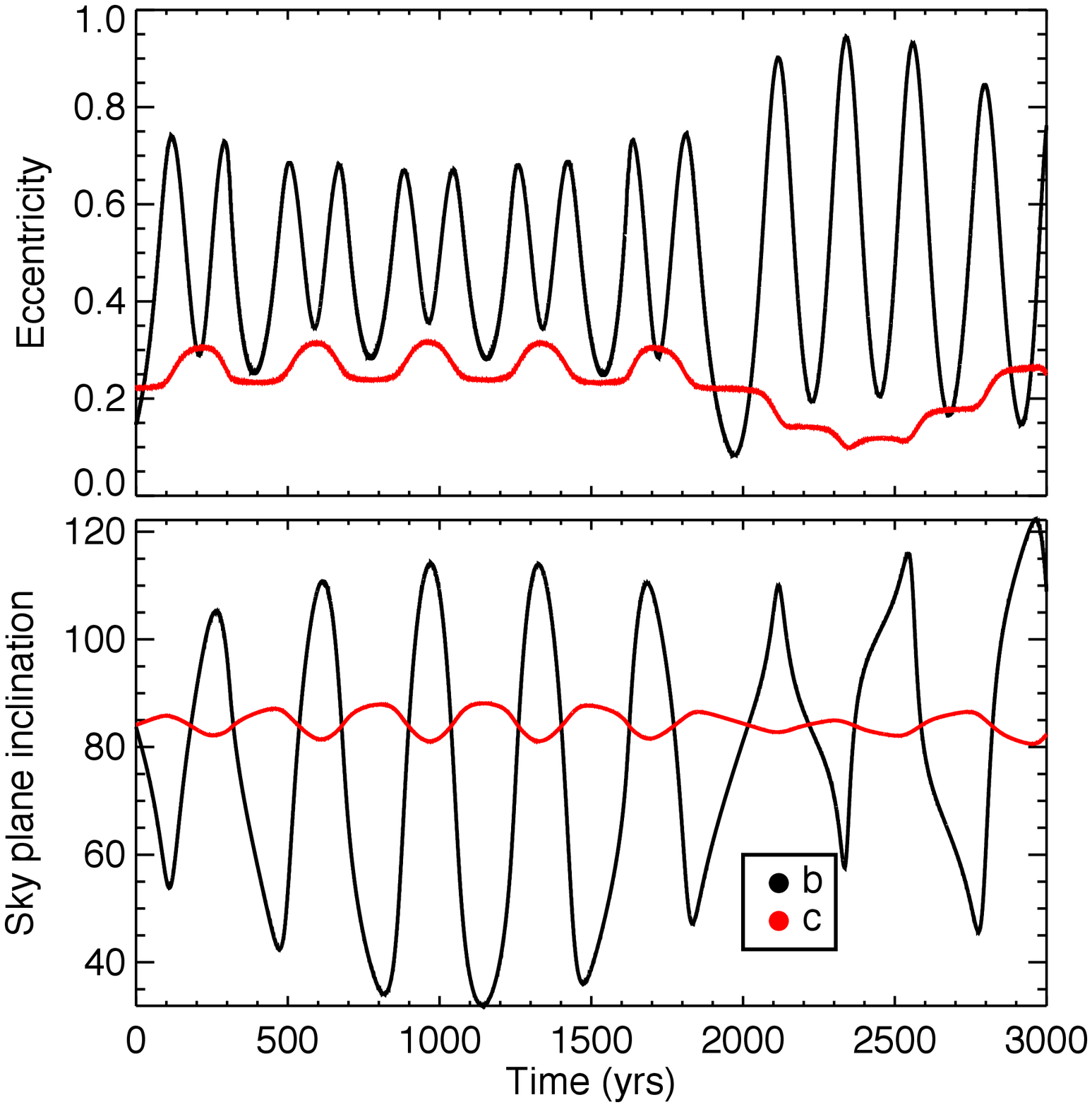}
\caption{ The eccentricity and sky plane inclination evolution of both planets for a 
randomly chosen initial condition with high mutual inclination. }
\label{fig:ecc_ev}
\end{center}
\end{figure}
 
 \begin{figure}[t!]
\begin{center}
\includegraphics[width=4.5in]{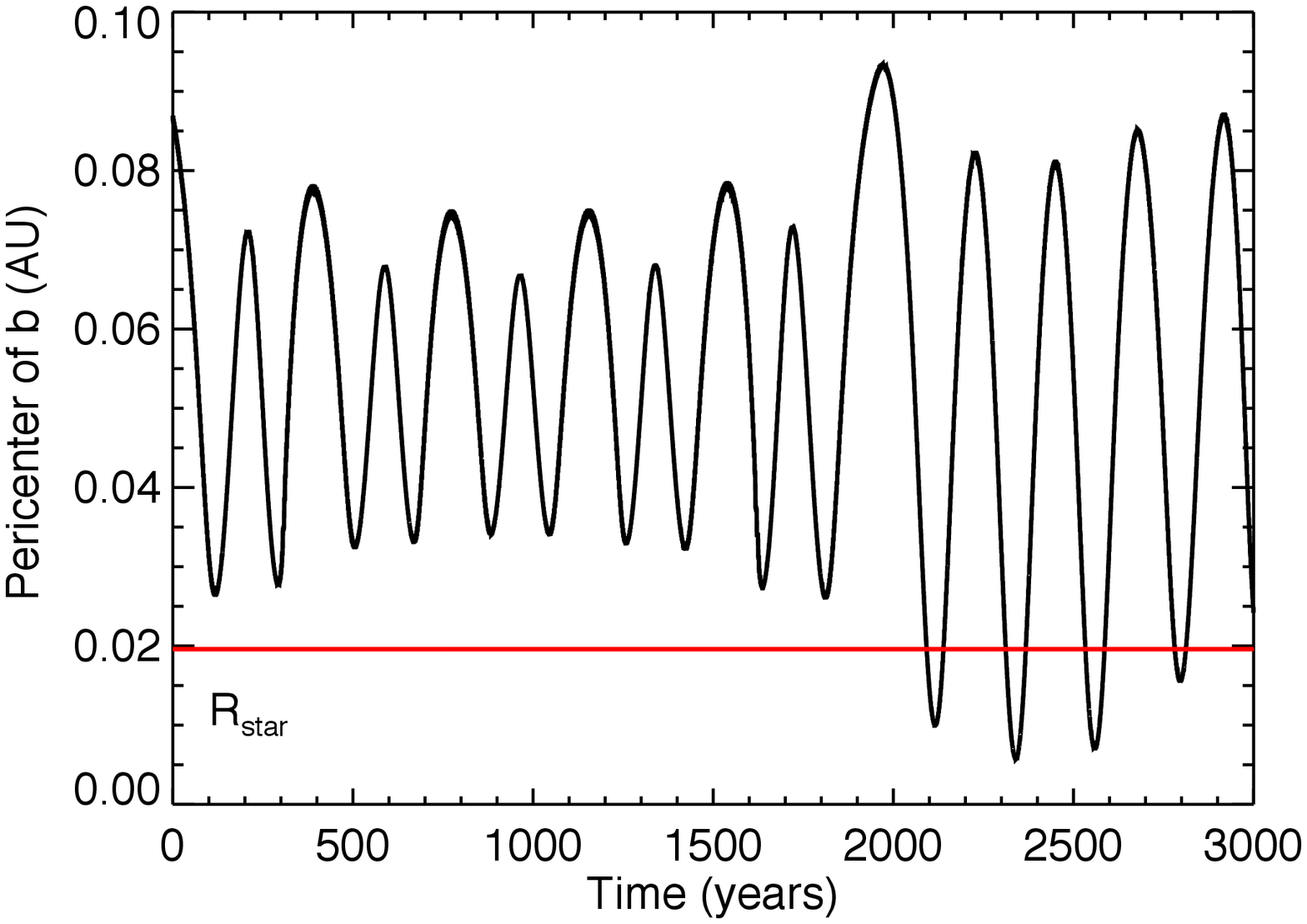}
\caption{The pericenter distance of the inner planet is show for the same initial 
condition as Figure \ref{fig:ecc_ev}. Note that since the bodies are treated as point 
particles the 
integrations do not stop when the collision actually occurs.}
\label{fig:peri}
\end{center}
\end{figure}
 
The timescale to reach this critical $e_b \approx 0.8$ is shorter for the initial conditions 
with the highest initial mutual inclination. Over $99\%$ of the initial conditions lead 
to a collision between the inner planet and the star within $10^4$ years. The remaining 
$1\%$ suffer collisions on timescales of $10^5$ years. The distribution of the time 
required before the inner planet crashes into the star is shown in Figure \ref{fig:dist}.
 
\begin{figure}
\begin{center}
\includegraphics[width=4.5in]{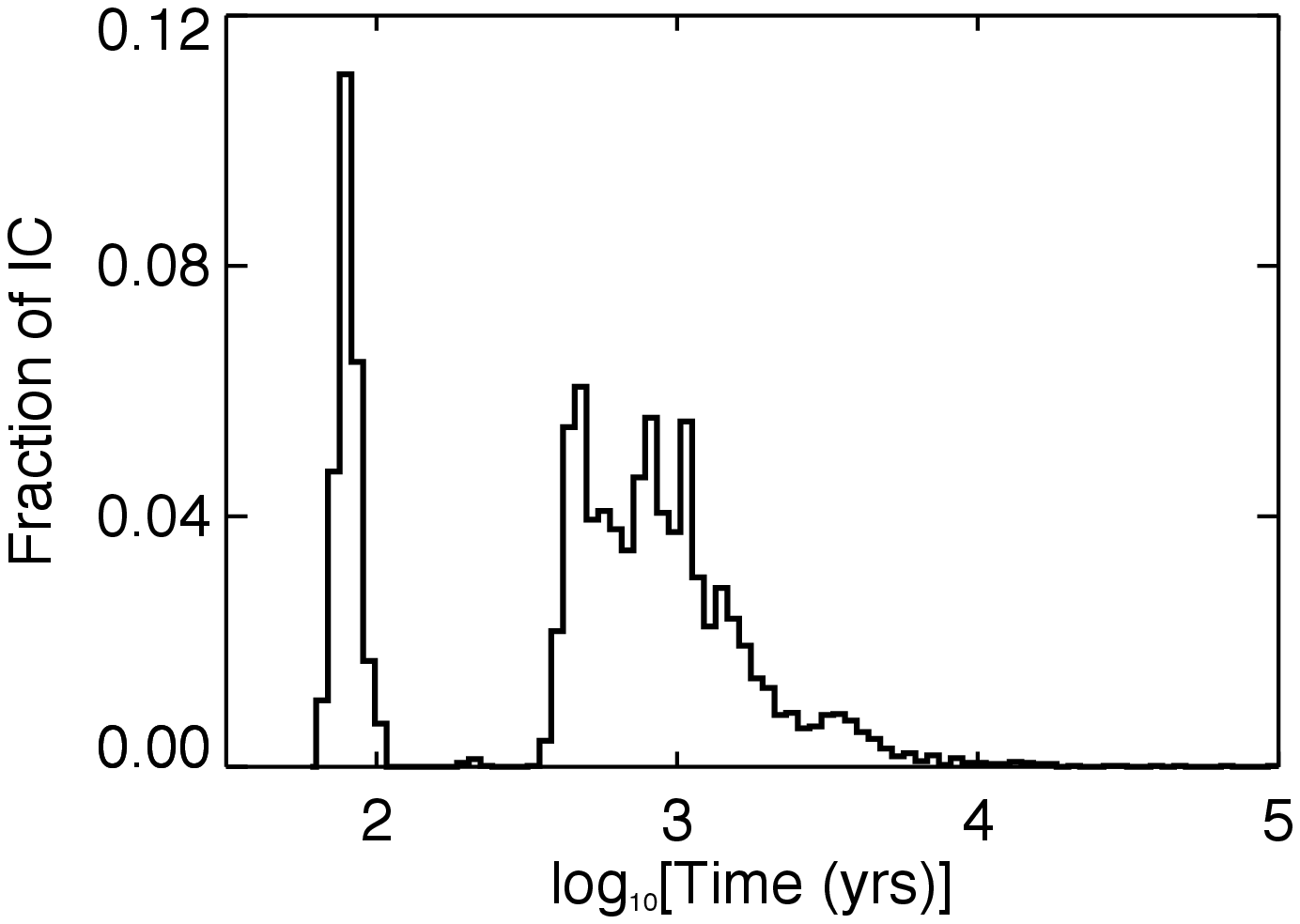}
\caption{The distribution of time required, in years, for the inner planet to reach a 
pericenter distance inside the surface of the star. }
\label{fig:dist}
\end{center}
\end{figure}

The typical timescales for collisions are very short, and hence tidal damping of 
eccentricity (or any other dissipative process) would not save the inner planet from 
this fate unless that dissipation was exceedingly efficient. Additionally, the 
dynamical effects of a far away perturber would likely be too weak to affect this 
result, and so we can conclude that the highly mutually inclined orbits which fit the data are 
unphysical.   
 
\subsection{Orbital Solutions with Low Mutual Inclinations}

We also studied the short-term stability of solutions with a low
initial mutual inclination. Out of $10^4$ solutions drawn from the
posterior distribution, two failed the Hill criterion. A set of 995
representative low-inclination initial conditions were integrated for
$5\times10^5$ years, or $\sim 2\times10^7$ orbits of the inner planet,
using a Wisdom-Holman symplectic integrator \cite{WH}. Symplectic
correctors were implemented to improve the accuracy of the
integrations \cite{Touma, Mercury}.  For these integrations, a
relativistic contribution was included because the typical precession
rate for the inner planet is on the order of $10^5$ years. We used a
dipole-like potential, which is straightforward to incorporate into
the symplectic integrator, to mimic the effect of general relativity
in the weak limit \cite{Nobili}. Energy was conserved to within one
part in $10^{10}$. For each of these initial conditions, we kept track
of the maximum, minimum and average semi-major axes, eccentricities,
pericenters, and sky-plane inclinations of the planets, as well as the
mutual inclination of the orbits. The evolution of the eccentricities and sky-plane 
inclinations for a typical initial condition is shown in Figure \ref{fig:ecc_ev_lowi}. 
The maximum eccentricity reached by
the inner planet in any of the integrations was 0.098. Note that the initial eccentricities 
no longer preferentially correspond to extrema in the secular cycle as was the case for the 
set of initial conditions with high mutual inclinations. Again, the semi-major 
axes were approximately constant ($\approx$0.05\% variation in
$a_c$ from the time-averaged value, and $\approx$0.3\% variation in
$a_b$), implying that the inner planet no longer crashes into or even
closely approaches the star. The typical pericenter evolution of the inner planet, 
relative to the radius of the star, is shown in Figure \ref{fig:peri_lowi}. The outer planet 
had a maximum
eccentricity of 0.107, and consequently the orbits do not come close
to crossing. 

\begin{figure}[t!]
\begin{center}
\includegraphics[width=5in]{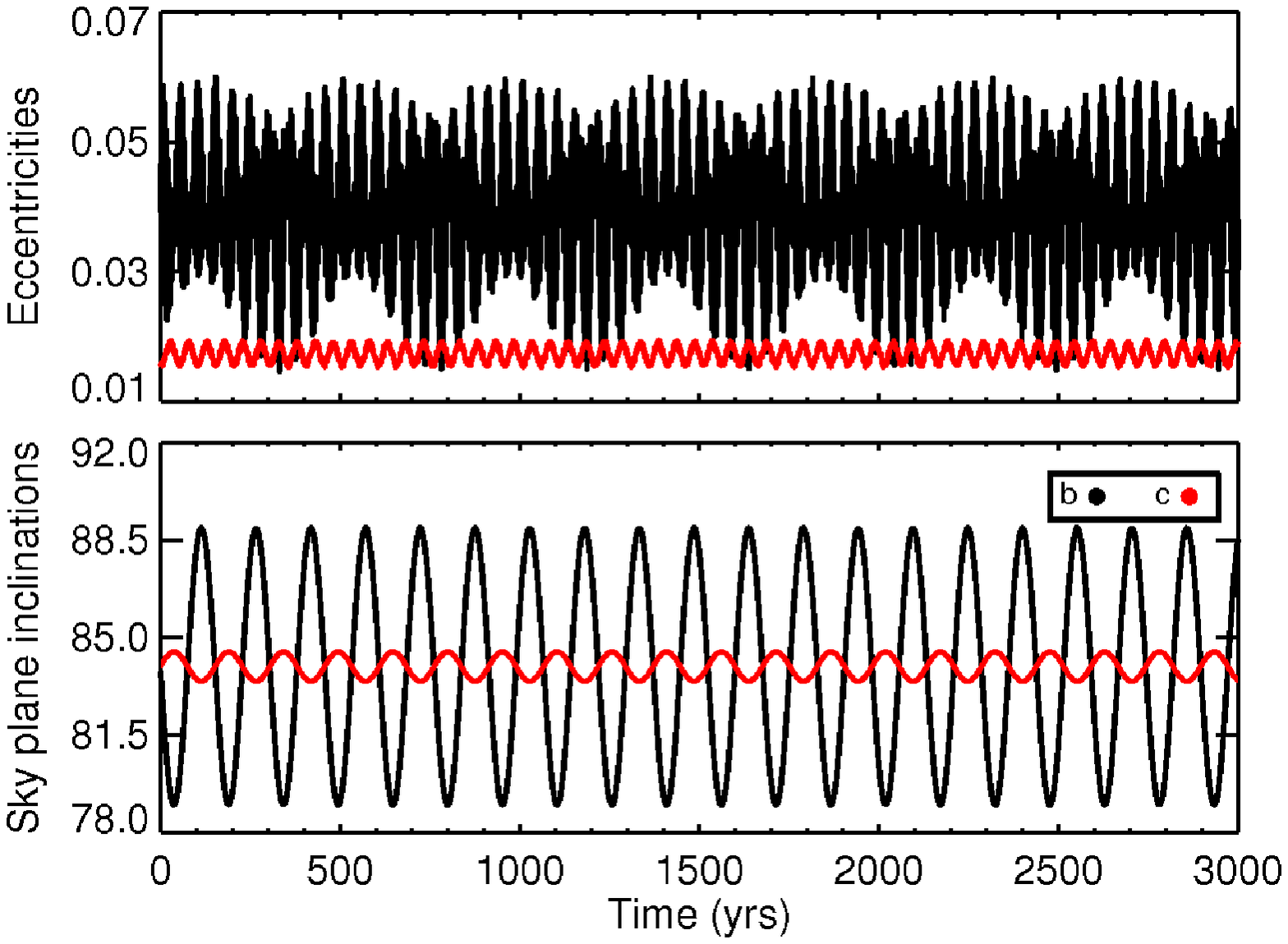}
\caption{ The eccentricity and sky plane inclination evolution of both planets for a 
typical initial condition with low mutual inclination ($I \sim 5^{\circ}$). }
\label{fig:ecc_ev_lowi}
\end{center}
\end{figure}
	
\begin{figure}[t!]
\begin{center}
\includegraphics[width=4.5in]{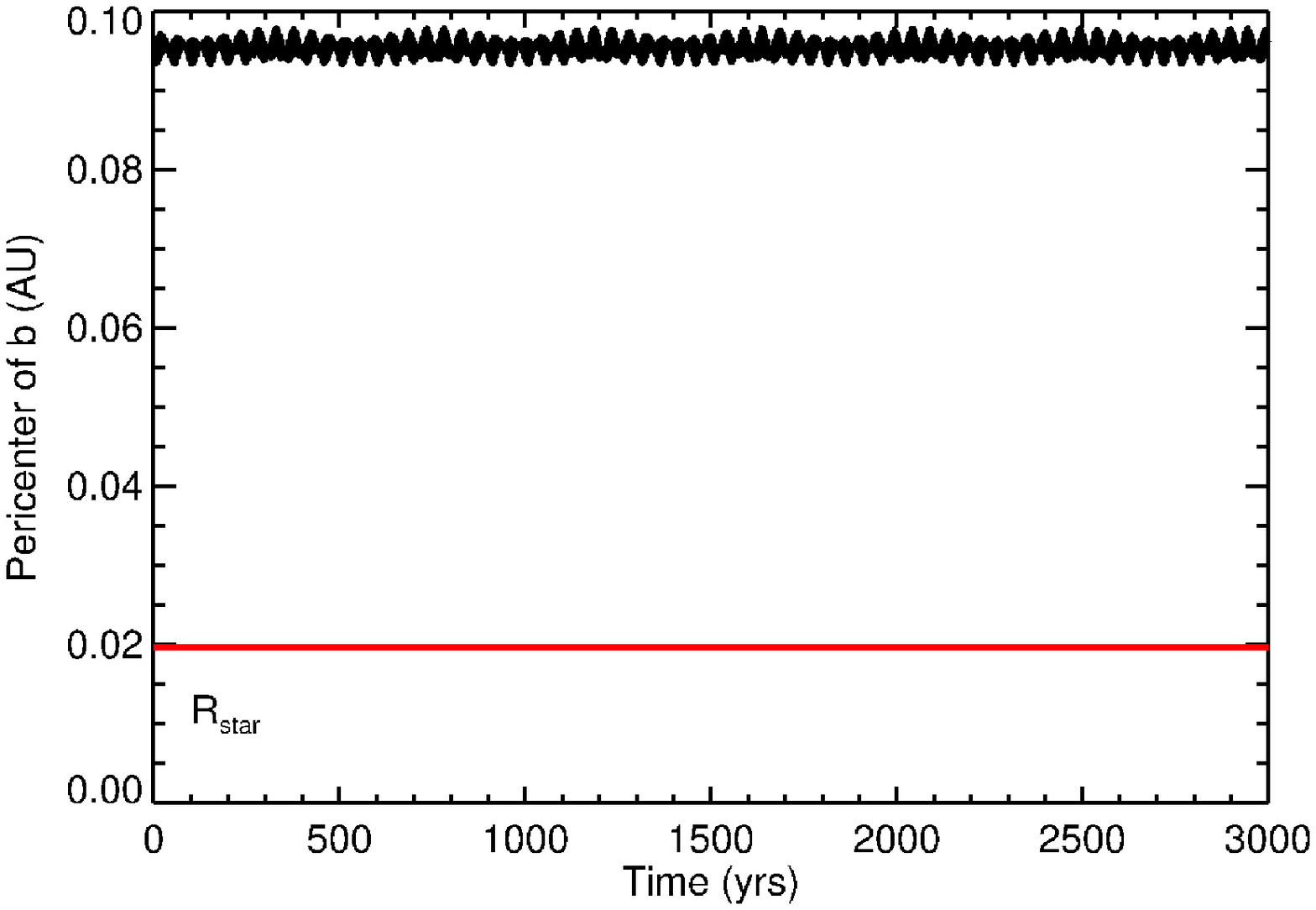}
\caption{The pericenter distance of the inner planet is show for the same initial 
condition as Figure \ref{fig:ecc_ev_lowi}. Initial conditions with low mutual inclinations 
do not lead to close approaches between the inner planet and the star.}
\label{fig:peri_lowi}
\end{center}
\end{figure}
 
With approximately half of the initial condition studied,
corresponding to those orbits with initial mutual inclinations $>$5
degrees, the inner planet periodically stops transiting. The typical
timescale for this behavior is 60,000 days (150 years). Similarly, the
outer planet stops transiting for about one quarter of the initial
conditions, with the same timescale. However, the mean inclinations of
the planets over the $5\times10^5$ years always corresponded to impact
parameters less than unity. For each initial condition, we monitored the time evolution 
of the planetary impact parameters. We found that there were 16 initial conditions where 
only planet c stopped transiting, 281 where only planet b stopped transiting, and 239 where 
both stopped transiting periodically. This analysis allows us to determine the probability 
of observing the Kepler 56 system as a different multiplicity system. For example, the 
probability of seeing the Kepler 56 system as an $n$-transiting system, where $n=0,1,$ 
or $2$, is $P_n  = \sum_{\rm all IC} P(IC) f(IC)_{t,n}$, where $P(IC)$ is the 
probability of the initial condition and $f(IC)_{t,n}$ is the fraction of the integration 
time that the initial condition spent in an $n$-transiting configuration. Since these 
initial conditions are drawn from the probability posterior distribution generated from 
fitting the data, the probability of each initial condition is already accounted for, and 
so we only need to set $P(IC) = 1/$(number of IC) as a normalization.  

Of the 239 initial conditions where both planets periodically stopped transiting, only 
one corresponded to the case where both planets were not in a transiting configuration 
with respect to our line of sight at the same time. The probability of seeing Kepler 56 
in the state where no planet is transiting is $P_0 = \frac{1}{995}\times$ the fraction 
of time this initial condition spent in the non-transiting configuration. We determined 
that $P_0 \approx 10^{-7}$. The probability of seeing the Kepler 56 system as a single 
transiting system is then approximately the sum over the remaining 994 initial conditions 
of the fraction of integration time that each planet is not in a transiting configuration, 
divided by 994. We found that $P_1 = 0.081$, and hence the probability of seeing the 
Kepler 56 system as double transiting system is $P_2  =0.92$. 

Finally, the mutual inclination remained essentially constant
throughout the integrations, without any long-term trend: these
initial conditions, identified by their low current mutual
inclinations, remain at a low mutual inclination. The typical
difference between the maximum and minimum mutual inclinations reached
over the course of the integrations was about 0.06 degrees.

These investigations are sufficient for the present purpose of
establishing the dynamical instability of the highly mutually inclined
solutions, and the plausible dynamical stability of the
low-inclination solutions.  A longer-term stability analysis
(integrations longer than $\sim 10^7$ years) would need to take into
account the effects of stellar (and tidal) evolution on the orbits.

\section{Specific Realization of the Dynamical Tilting Hypothesis}

The radial velocity data show a long-term drift due to a third companion in the Kepler-56 
system. Based on the linear trend of $0.8$\,m\,s$^{-1}$\,d$^{-1}$ determined from the 
simultaneous fit of the Kepler and radial-velocity data 
(\S\ref{sec:photodyn}), a third body in 
a circular orbit with a period $P_{\rm d}$ would have a minimum mass of 
$1.6 M_{\rm J} (P_{\rm d}/\rm{yr})^{4/3}$. Hence, if the long-term velocity drift is 
seen to halt and reverse direction within
the next few years, the third body would be implicated as another planet. The orbit 
of this more massive, outer companion dominates the angular momentum of the system, and hence 
can have a strong influence on the spin-orbit angle of the inner planets.  
If the third body is inclined by an angle $I_3$ relative to the mean
orbital plane of the transiting planets, 
it would produce a torque on the planes of the inner planets and cause
them to precess cyclically around the total 
angular momentum with a maximum inclination of $2I_3$. 
The planets in turn also cause the star to precess around the total angular momentum. 
However, the star precesses on a much slower timescale, 
resulting in large inclinations between the spin-axis of the host star and the 
orbital plane of the inner planets. The mutual inclination of the inner planets, on the 
other hand, remains low due to their compact orbits \cite{takeda08}. 
Figure \ref{fig:sketch} shows 
a graphical representation of this dynamical tilting hypothesis.

To quantify this scenario, 
we used the direct 3-planet code of Mardling \& Lin \cite{mardling02} to 
integrate the motion of such a system 
explicitly.  These equations track the orbital trajectories of three planets, as well 
as the spin rate and direction of the star and the innermost
planet. For concreteness we adopted
the current star's parameters, and assumed
the outer body has a mass of 3.3~$M_{\rm J}$ 
planet with a semi-major axis of 2 AU, and eccentricity of 0.4. We furthermore assume a 
mutual inclination of
$25^\circ$ of the outer planet with respect to the inner planet, the middle planet, and 
the stellar equatorial
plane, all of which are initially aligned with one another. 
This choice for the inclination of the third body 
corresponds to a typical value for planets on wide orbits produced by 
planet-planet scattering \cite{chatterjee08,nagasawa08}.
Table \ref{tab:ini} gives the initial conditions of the simulation. 
Figure \ref{fig:spin} shows the evolution of the inclinations of all relevant 
angles of the system over a timescale of $3\times10^5$ years. 
As suggested by the qualitative discussion above, the 
spin-orbit angle between the host star and the inner, transiting system can reach 
large angles due to the outer planet on an inclined, eccentric orbit, consistent 
with our observations in the \id\ system.

This simulation provides a proof of concept for the scenario described
in the main text. We emphasize, though, that the properties of the
companion were chosen somewhat arbitrarily.  Furthermore, for this
calculation we adopted the properties for the host star in its current
evolutionary stage, but it should be borne in mind that the dynamical
action may have occurred long ago when the star was on the main
sequence or even the pre-main-sequence. 
Dynamical simulations including the evolution of the host star would allow
study the effects of mass loss or tides on the planets \cite{voyatzis13}, and  
potentially constrain when the instability occurred.
However, for a smaller host star, the
precession of the star would occur with an even longer period, and
hence produce a smaller torque on the inner planets than it currently
does.  Hence there would be even less difficulty producing a large
spin-orbit misalignment.

 \begin{table}
\centering
\begin{tabular}{l c}
\textit{Host Star} \\
\hline
Mass (\msun)								& 1.32			\\
Radius (\rsun)								& 4.23			\\
Apsidal motion constant 					& 0.004         \\
Moment of inertia coefficent 			    & 0.02          \\
Obliquity (deg)								& 0.0			\\ 
\textit{Planet b} \\
\hline
Eccentricity 								& 0.0       	\\
Semi-major axis (AU) 						&	0.1028      \\        
Inclination relative to third companion (deg) 	& 25.0 			\\  
Argument of perigee  (deg)					& 0.0           \\  
Longitude of line of nodes  (deg)		    & 0.0           \\        
True anomaly (deg)								& 57.0			\\    
Mass ($M_{\rm J}$)							& 0.069         \\
Radius ($R_{\rm J}$)						& 0.3			\\      
Apsidal motion constant 					& 0.15          \\         
Moment of inertia coefficient 			    & 0.25          \\
\textit{Planet c} \\
\hline
Eccentricity 								& 0.0       	\\
Semi-major axis (AU)						& 0.1652      	\\        
Inclination relative to third companion (deg) 	& 25.0 			\\  
Argument of perigee  (deg)					& 0.0           \\  
Longitude of line of nodes  (deg)		    & 0.0           \\        
True anomaly (deg)							& 182.0 		\\    
Mass ($M_{\rm J}$)							& 0.569         \\
\textit{Third Companion} \\
\hline
Eccentricity 								& 0.4       	\\
Semi-major axis (AU) 						& 2.0      		\\        
True anomaly (deg) 							& 256.0 		\\    
Mass ($M_{\rm J}$)							& 3.3         	\\
\end{tabular}
\caption{Initial conditions of the dynamical simulation of the \id\ system, 
including a third companion on an eccentric and inclined orbit.}
\label{tab:ini}
\end{table}

\begin{figure}
\begin{center}
\includegraphics[width=\hsize]{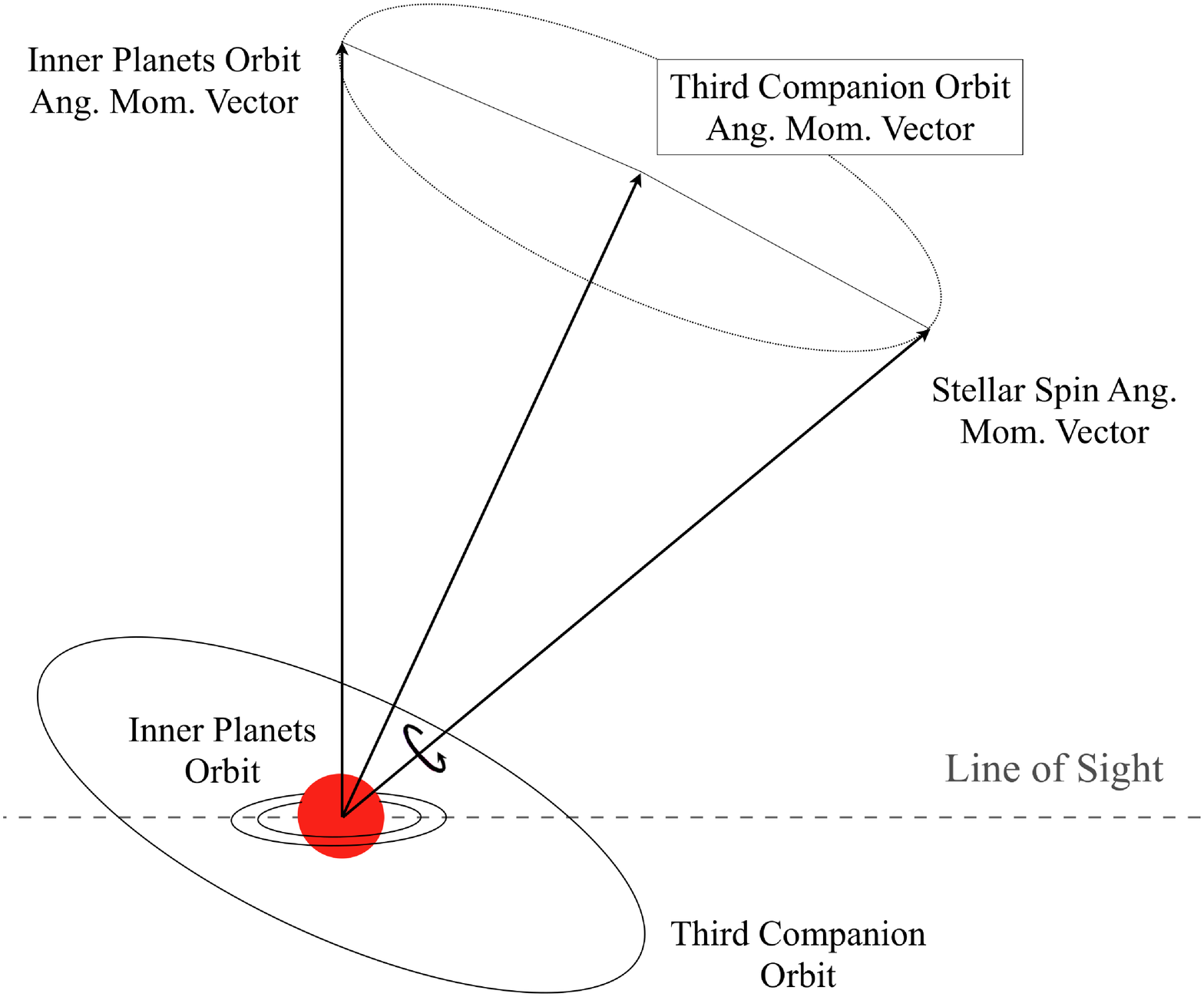}
\caption{Graphical illustration of the dynamical tilting hypothesis for 
the \id\ system. Note that the sizes are not to scale.}
\label{fig:sketch}
\end{center}
\end{figure}

\begin{figure}
\begin{center}
\includegraphics[width=12.25cm]{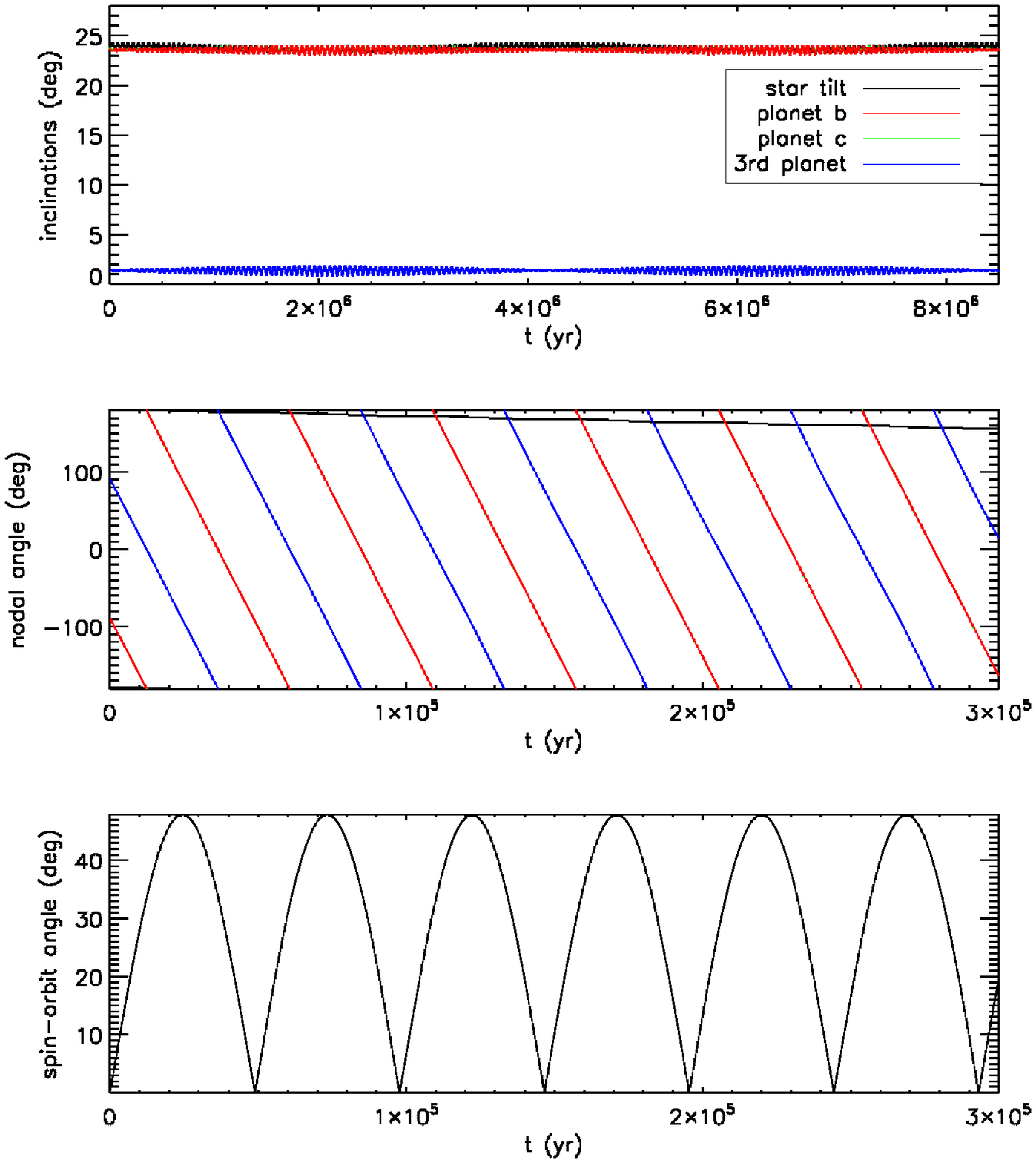}
\caption{Simulation results demonstrating the generation of
spin-orbit misalignment from the torque from a wide-orbiting, inclined companion. Initially, 
the observed system of two transiting planets is placed on coplanar orbits in the equatorial 
plane of their host star. A massive, inclined, and eccentric planet is placed exterior 
to them.  Panel (a): Inclinations with respect to the Laplace plane (the plane normal 
to the total angular momentum), each of which stay nearly constant.  
Panel (b): Nodal angles versus simulation time. The inner and middle planets precess 
in concert, with the same nodal 
angle. The outer planet precesses at the same rate, but $180^\circ$ out of phase. The star 
precesses very slowly, due to its weak coupling to the planets.  Panel (c): 
Angle between the stellar equator plane and the inner planet's orbital plane.
Both planets and the star remain $24^\circ$ inclined from the Laplace 
plane, but they precess at different rates, and hence are periodically misaligned by 
$48^\circ$.}
\label{fig:spin}
\end{center}
\end{figure}

\newpage

\bibliography{/Users/daniel/science/codes/latex/references}

\begin{thebibliography}{100}
\bibitem{SOM}
Materials, methods are available as supporting material~on Science~Online. .
\end{thebibliography}

\bibliographystyle{Science}

\end{normalsize}

\end{document}